\documentclass[journal=nalefd,manuscript=letter]{achemso} 
\setkeys{acs}{etalmode=truncate,maxauthors=10}

\usepackage{graphpap}
\usepackage{color}
\usepackage[normalem]{ulem}
\usepackage{ulem}
\usepackage{amsmath}
\usepackage[T1]{fontenc}
\usepackage[latin9]{inputenc}
\usepackage{array}
\usepackage{multirow}
\usepackage{amssymb}
\usepackage{textcomp}
\usepackage{soul}

\usepackage{graphicx}
\usepackage{braket}
\usepackage{amsmath}
\usepackage{bm}
\usepackage{caption}
\usepackage{subcaption}
\usepackage[section]{placeins}
\usepackage[english]{babel}
\usepackage{natbib}
\usepackage{xspace}

\newcommand{\kb}{\mathbf{k}}

\newcommand{\eck}{\epsilon_{c\kb}}
\newcommand{\evk}{\epsilon_{v\kb}}

\newcommand{\es}{\epsilon_s}
\newcommand{\mote}{MoTe$_2$\xspace}
\newcommand{\abinitio}{\textit{ab initio}\xspace}
\newcommand{\apr}{$A_1^\prime$\xspace}
\newcommand{\apa}{$A_1^\prime$(a)\xspace}

\newcommand{\apc}{$A_1^\prime$(b)\xspace}
\newcommand{\epr}{$E^\prime$\xspace}

\title{Quantum interference effects in resonant Raman spectroscopy of single- and triple-layer MoTe$_2$ from first principles}

\author{Henrique P. C. Miranda}
\affiliation{Physics and Materials Science Research Unit, University of Luxembourg, 162a avenue de la Fa\"iencerie, L-1511 Luxembourg, Luxembourg, EU}
\email{henrique.pereira@uni.lu}

\author{Sven Reichardt}
\affiliation{Physics and Materials Science Research Unit, University of Luxembourg, 162a avenue de la Fa\"iencerie, L-1511 Luxembourg, Luxembourg, EU}
\alsoaffiliation{JARA-FIT and 2nd Institute of Physics, Otto-Blumenthal-Stra{\ss}e, 52074 Aachen, Germany, EU}

\author{Guillaume Froehlicher}
\affiliation{Universit\'e de Strasbourg, CNRS, Institut de Physique et Chimie
des Mat\'eriaux de Strasbourg (IPCMS), UMR 7504, F-67000 Strasbourg, France, EU}

\author{Alejandro Molina-S\'anchez}
\affiliation{Physics and Materials Science Research Unit, University of Luxembourg, 162a avenue de la Fa\"iencerie, L-1511 Luxembourg, Luxembourg, EU}

\author{St\'ephane Berciaud}
\affiliation{Universit\'e de Strasbourg, CNRS, Institut de Physique et Chimie
des Mat\'eriaux de Strasbourg (IPCMS), UMR 7504, F-67000 Strasbourg, France, EU}

\author{Ludger Wirtz}
\affiliation{Physics and Materials Science Research Unit, University of Luxembourg, 162a avenue de la Fa\"iencerie, L-1511 Luxembourg, Luxembourg, EU}

\date{\today}

\begin{document}

\begin{abstract}
We present a combined experimental and theoretical study of resonant Raman spectroscopy in single- and triple-layer MoTe$_2$.
Raman intensities are computed entirely from first principles by calculating finite differences of the dielectric susceptibility.
In our analysis, we investigate the role of quantum interference effects and the electron-phonon coupling.
With this method, we explain the experimentally observed intensity inversion of the $A^\prime_1$ vibrational modes in triple-layer MoTe2 with increasing laser photon energy.
Finally, we show that a quantitative comparison with experimental data requires the proper inclusion of excitonic effects.
\end{abstract}

\maketitle

\section{Introduction}
Transition metal dichalcogenides (TMDs) are good candidates for nanoengineering due to their quasi-two-dimensional nature.
The weak interlayer interaction allows the fine-tuning of the electronic and vibrational properties of the nanostructure by stacking different types and numbers of layers.\cite{geim_van_2013}

To characterize the properties of these nanostructures, Raman spectroscopy is a useful and accurate technique, which simultaneously probes their vibrational and optical properties.
It yields information about the lattice symmetry, the vibrational eigenmodes, and optically active electronic transitions, including excitonic effects.\cite{cardona_light_scattering_solids_II,loudon_theory_1963}

In particular, when in the resonant regime, the Raman intensities show a strong dependence on the laser photon energy for certain phonon modes, as was shown for 
MoSe$_2$,\cite{soubelet_resonance_2016,kim_davydov_2016} MoS$_2$,\cite{lee_anomalous_2015} and WS$_2$.\cite{staiger_splitting_2015}
This dependence allows the identification of excitonic states and the investigation of their coupling to phonons, as demonstrated for MoS$_2$,\cite{carvalho2015,scheuschner_interlayer_2015} WS$_2$, and WSe$_2$.\cite{del_corro_atypical_2016}
In \mote, measurements also show such a strong dependence.\cite{yamamoto_strong_2014,ruppert_optical_2014,froehlicher2015,grzeszczyk_raman_2016,song_physical_2016}
In the case of triple-layer \mote, it was observed that the intensity ratio between the lowest- and highest-frequency modes belonging to the same Davydov triplet significantly changes with laser photon energy.\cite{froehlicher2015,grzeszczyk_raman_2016,song_physical_2016}
The change of the Raman intensities with laser photon energy in \mote and in TMDs in general is yet to be fully understood and few \abinitio studies are present in the literature.\cite{guo_double_2015}
More recently, the experimental observation of the temperature dependence of the Raman intensities was reported.\cite{golasa_resonant_2017}
Single-layer MoTe$_2$ is a near-infrared (1.1~eV at room temperature) direct optical band gap semiconductor, as such it is possible
to probe excitonic states with visible photon energies.\cite{ruppert_optical_2014,froehlicher_direct_2016}
Additionally the Davydov split modes appear prominently at visible (hence easily available) laser photon energies.\cite{froehlicher2015,grzeszczyk_raman_2016,song_physical_2016}

In this work, we explain the dependence of the one-phonon Raman intensities on the laser photon energy using computational simulations and compare them with experimental results.
The accurate description of resonant Raman scattering is challenging due to the interplay between electronic correlation and electron-phonon coupling.
Up to now, most theoretical studies have focused on the non-resonant regime using simpler models like the bond-polarizability model or density functional perturbation theory.\cite{luo_anomalous_2013,luo_effects_2013,umari_raman_2001}
However, these methods assume static electromagnetic fields, which is not applicable in the resonant case where the \emph{dynamic} dielectric response needs to be accounted for.
Resonant Raman spectroscopy has also been studied using empirical models fitted from experiments to describe the electronic bands, phonon dispersion and electron-phonon coupling.\cite{cantarero_excitons_1989}
More recently, a study on the double-resonant Raman process in \mote investigated the resonance surface using calculations of the electronic structure and phonon dispersion.\cite{guo_double_2015}

Here we use an \textit{ab initio} approach to calculate the first-order Raman susceptibility as a function of laser photon energy.
We calculate the Raman susceptibility by approximating the derivative of the dielectric response function with respect to lattice displacements with finite differences.\cite{gillet2013,del_corro_atypical_2016}
To this end, we combine different \textit{ab initio} methods:
we calculate the ground state properties using density functional theory (DFT), the phonons with density functional perturbation theory (DFPT), and the optical absorption spectra both in the independent-particle approximation and including many-body effects.
We discuss the main qualitative features on the independent-particle level and show that the inclusion of excitonic effects provides a reliable quantitative description of the Raman spectrum, in very good agreement with experimental results.
Moreover, the calculations reproduce the experimentally reported\cite{froehlicher2015,grzeszczyk_raman_2016,song_physical_2016} dependence of the intensity ratio of the $A^\prime_1$ Davydov triplet as a function of laser photon energy. Finally we give an explanation of the results in terms of quantum interference effects.

\section{Raman intensities from first principles}\label{sec:raman}

The experimental observable of interest, the Raman intensity, is, in the case of phonon emission (Stokes scattering), given by\cite{loudon_theory_1963,birman_theory_1966,cardona_light_scattering_solids_II}
\begin{align}
I (\omega_L) \propto \sum_\mu (\omega_{\rm L}-\omega_\mu)^4 |(\vec{e}_{\rm S})^\dagger \bm{\alpha}_\mu (\omega_L) (\vec{e}_{\rm L})|^2\frac{n_\mu+1}{2\omega_\mu}.
\label{eqI}
\end{align}
Here, $\bm{\alpha}_\mu(\omega)$ is the Raman susceptibility tensor, $\vec{e}_{\rm L}$ and $\vec{e}_{\rm S}$ are the polarization vectors of the incoming and scattered light, respectively, $\omega_{\rm L}$ is the frequency of the incoming light, $\omega_{\mu}$ denotes the frequency of phonon mode $\mu$, and $n_\mu$ represents its occupation factor.
In the frozen-phonon limit, the Raman tensor equals the change of the dielectric susceptibility $\bm{\chi}(\omega)$ with atomic displacements\cite{cardona_light_scattering_solids_II}
\begin{align}
\bm{\alpha}_\mu(\omega) = \sum_{\tau, i} \frac{\partial \bm{\chi}(\omega)}{\partial R_{\tau, i}} Q^{\tau,i}_\mu, \label{eq:raman_tensor}
\end{align}
where $R_{\tau,i}$ is the position of atom $\tau$ in the Cartesian direction $i$, and $Q_\mu$ the eigenvector of the phonon mode $\mu$, normalized according to
\begin{align}
\sum_{\tau, i} M_{\tau} Q^{\tau, i}_\mu Q^{\tau, i}_\nu = \delta_{\mu\nu}.
\end{align}
where $M_{\tau}$ denotes the mass of atom $\tau$.
This formulation allows us to account for many-body effects in the Raman susceptibility by incorporating them in the calculation of the dielectric response.
At this level, different well-tested implementations are available in a fully \textit{ab initio} framework which allow the inclusion of excitonic and electronic correlation effects, which are especially relevant in TMDs.\cite{qiu_optical_2013,molina-sanchez_effect_2013}

The frozen-phonon approximation is valid at energies that fulfill the condition
\begin{align}
\hbar\omega_{\mathrm{\mu}} \ll \left| \hbar\omega_{\rm L} - \Delta E + i\gamma \right|, 
\end{align}
where $\Delta E$ represents the energy of an electronic transition, $\hbar\omega_{\rm L} = E_{\rm L}$ is the photon energy of the incoming laser light (from now on designated simply as laser energy) and $\gamma$ is the broadening, i.e., the inverse lifetime, of the electronic excitation. 
In the non-resonant regime, $E_{\rm L}$ is far away from any electronic transition energy and this condition is automatically satisfied.
In the resonant regime, where the laser energy always matches the energy of an electronic transition, the relevant condition is that the phonon energy ($\sim$20-25~meV) is smaller than the electronic broadening.
At room temperature the broadening due to electron-phonon coupling is around 100 meV\cite{molina-sanchez_temperature-dependent_2016} and therefore the frozen-phonon approximation is reasonable.

This approach explicitly captures the laser-energy dependence inherent to the Raman susceptibility tensor, which is crucial for studying resonance effects.
This formulation goes beyond the bond polarizability model and DFPT, which assume static electromagnetic fields, and are therefore only valid in the non-resonant regime.\cite{lazzeri_first-principles_2003,veithen_nonlinear_2005}

\subsection{Electronic structure and phonons}\label{sec:gs}
The electronic structure of \mote is calculated using DFT within the local density approximation (LDA), as implemented in the PWscf code of the Quantum ESPRESSO suite.\cite{giannozzi_quantum_2009}
We include the semi-core 4s and 4p states in the pseudopotential of molybdenum and account for spin-orbit interaction by employing spinorial wave functions.
The charge density is calculated using a plane-wave energy cutoff of 100~Ry and a $16\times 16\times 1$ $\bf k$-point grid for both the single- and triple-layer calculation.
For the lattice parameter, we use the experimental value of 3.52~\AA. \cite{podberezskaya_crystal}

\begin{table}
\center
\label{table:sl}
\begin{tabular}{lccccc}
    Mode & $A{^\prime_1}$ &  $E^{\prime}(x)$ & $E^{\prime}(y)$\\\hline
    Raman tensor $\bm{\alpha}_\mu$ &
    $\begin{bmatrix}
    a &   &  \\
      & a &  \\
      &   & b\\
    \end{bmatrix}$&
    $\begin{bmatrix}
       &  d &    \\
     d &    &    \\
       &    &    \\
    \end{bmatrix}$&
    $\begin{bmatrix}
     d &    &    \\
       & -d &    \\
       &    &    \\
    \end{bmatrix}$\\\hline
    Single-layer      &\phantom{(a)} 174.6          (171.5)  & \multicolumn{2}{c}{ 238.3 (236.5)           } \\
    Freq. (cm$^{-1}$) &                                      &        \\\hline
    Triple-layer      &          (a) 173.6          (169.4)  & \multicolumn{2}{c}{ 235.5 (234.7)           } \\
    Freq. (cm$^{-1}$) &\phantom{(a)} 175.1 \phantom{(000.0)} & \multicolumn{2}{c}{ 238.0 \phantom{(000.0)} } \\
                      &          (b) 176.4          (172.6)  & \multicolumn{2}{c}{ 239.0 (234.7)           } \\
\end{tabular}
\caption{
Calculated and experimental\cite{froehlicher2015} (in parentheses) phonon mode frequencies and corresponding form of the Raman tensor for the space group D$_{\rm 3h}$.\cite{loudon_theory_1963}
We distinguish the two Raman active $A^\prime_1$ modes in triple-layer \mote using the letters (a) and (b).
The triple-layer mode with frequency 175.1 cm$^{-1}$ is Raman inactive and belongs to the $A^{\prime\prime}_2$ representation.
All other listed modes are Raman-active.
The calculated splitting of the $E^\prime$ mode in triple-layer \mote is not observed experimentally.
This mode, however, is not studied in detail here. For a complete discussion see Ref.\citenum{froehlicher2015}.
}\label{tab:phonon_modes}
\end{table}

The phonons of \mote are calculated using DFPT.
Due to momentum conservation, only phonon modes at $\Gamma$ participate in first-order Raman scattering, as the magnitude of the light momentum is negligible compared to the crystal momentum.
The Raman-active phonon modes of interest are reported in Table~\ref{tab:phonon_modes}.
Both single- and triple-layer \mote belong to the space group D$_{\rm 3h}$.
We refer to the different phonon modes by their irreducible representation label in the Mulliken notation.
The phonon modes of single-layer \mote are denoted by $E^\prime$ and $A^\prime_1$ for the in-plane and out-of-plane modes, respectively.\cite{molina-sanchez_vibrational_2015}
When going from single-layer to triple-layer \mote, the $A^\prime_1$ mode splits into a Davydov triplet composed of two Raman-active $A^\prime_1$ modes, which we denote by $A^\prime_1$(a) and $A^\prime_1$(b), and one IR-active $A^{\prime\prime}_2$ mode.
In this work, we will study the experimentally observed inversion of the Raman intensity ratio between the $A^\prime_1$(a) and (b) modes as a function of laser energy as shown in Figure~\ref{fig:Fig_IPCMS}.

\subsection{Optical absorption}\label{sec:abs}

We calculate the optical absorption on two levels of theory:
in the first approach, we treat electrons and holes as independent particles (IP), while in the second case, we include many-body effects due to electron-electron and electron-hole interaction perturbatively using the GW approximation and Bethe-Salpeter equation (BSE).\cite{onida_electronic_2002}

\subsubsection{Independent-particle approximation}

The expression for the IP dielectric susceptibility can be derived from time-dependent perturbation theory and is given by \cite{baroni_abinitio_1986}
\begin{align}
\chi^{ij}(\omega)
\propto \sum_{\kb vc} \left[ \frac{(\Lambda^i_{cv\kb})^* \Lambda^j_{cv\kb}}
{\hbar\omega - (\eck-\evk) + i\gamma} + (\omega \to -\omega) \right],
\label{eq:chi_ip}
\end{align}
where $\Lambda^i_{cv\kb} = \braket{\psi_{c\kb}| p^i/m_e |\psi_{v\kb}}$ denotes the electron-light coupling (ELC) matrix elements, also referred to as dipole matrix element, and $\evk$ and $\eck$ are the DFT valence and conduction bands energies, respectively.
The index $i$ denotes the Cartesian component of the ELC, while the parameter $\gamma$ represents the sum of the electron and hole broadening.
We use a constant broadening of 100~meV.
The calculation of the ELC was performed using the \texttt{yambo} code. \cite{marini_yambo:_2009}
The absorption spectrum is proportional to the imaginary part of the diagonal elements of the dielectric susceptibility tensor $\boldsymbol{\chi}(\omega)$.

\subsubsection{Many-body perturbation theory}

The two-dimensional character of \mote reduces the dielectric screening and hence many-body effects are more pronounced than in three-dimensional materials.
Such effects manifest themselves as significant corrections to the electronic band energies and in large excitonic binding energies.
We account for these effects by combining the GW method and the BSE.\cite{rohlfing_electron-hole_2000}
GW calculations were performed non-self-consistently (G$_0$W$_0$) using a $36\times 36\times 1$ sampling of the Brillouin zone (BZ) and a 40~Ry cutoff for the plane-wave basis.
A converged quasi-particle band gap was obtained by including 120 electronic bands for single- and 360 bands for triple-layer \mote.
It should be noted that an accurate GW correction requires the inclusion of the semi-core states in the Mo pseudopotential.\cite{rohlfing_quasiparticle_1995}
In order to avoid spurious interactions between periodic copies of the layers along the z-direction, we apply a Coulomb cutoff.\cite{rozzi_exact_2006} 

We account for electron-hole interactions by solving the BSE with a statically screened Coulomb potential.\cite{rohlfing_electron-hole_2000}
In terms of exciton energies $\es$, exciton-light coupling matrix elements $\Gamma^i_s$, and excitonic broadening $\gamma$, the dielectric susceptibility reads:
\begin{align}
\chi^{ij}(\omega) \propto
\sum_{s}
\frac{(\Gamma^i_{s})^* \Gamma^j_{s}}
{\hbar\omega - \es + i\gamma} + (\omega \to -\omega)\label{eq:chi_bse}.
\end{align}
The BSE calculations were performed using a 30~Ry cutoff for the plane-wave basis and a $36\times 36\times 1$ $\kb$-point grid to sample the Brillouin zone.
We include electronic transitions inside a 3~eV window (see Supporting Information for details of the GW and BSE calculations).

\subsection{Raman susceptibility tensor}

The Raman susceptibility tensor $\bm\alpha_\mu(\omega)$ of phonon mode $\mu$ is calculated by approximating the directional derivative of $\boldsymbol{\chi}(\omega)$ with the finite differences method.
For this, we evaluate the dielectric susceptibility at the two displaced positions $\vec{R}^\pm_\tau = \vec{R}_\tau \pm \delta \vec{Q}^\mu_\tau$ and divide by the amplitude of the two displacements.

An important practical drawback of this method is that the displacements according to certain phonon modes break some symmetries of the crystal.
This in turn increases the computational cost of the calculation with respect to the fully symmetric absorption calculation.
To reduce the computational cost, we extrapolate the GW correction from the undisplaced to the displaced case using a scissor operator, which incorporates the stretching of the bands.
This scissor operator is kept fixed for all calculations (see Supporting Information).
In addition, note that both the real and imaginary part of the dielectric susceptibility enter in the calculation of the Raman susceptibility.
The real part is known to converge more slowly with the number of bands.

In the IP picture we can further analyze the Raman susceptibility tensor by splitting it up into the contributions from individual $\kb$-points.
To this end, we note from Eq.~\ref{eq:chi_ip} that we can
represent the susceptibility $\chi^{ij}(\omega)$ as a sum over $\kb$:
\begin{equation}
  \chi^{ij}(\omega) = \sum_{\kb}\chi_\kb^{ij}(\omega),
\end{equation}  
where the term $\chi_\kb^{ij}(\omega)$ contains contributions from all electronic transitions at that $\kb$-point.
Analogously, we write the Raman susceptibility from Eq.~\ref{eq:raman_tensor} as $\alpha^{ij}(\omega) = \sum_\kb \alpha^{ij}_\kb(\omega)$. 

Contrary to the dielectric susceptibility, in which $\chi^{ij}(\omega)$ is the sum of all $\chi_{\kb}^{ij}(\omega)$, the Raman intensity is the \emph{square} of the sum of $\alpha^{ij}_\kb(\omega)$:
\begin{align}
\begin{split}
I \propto \left| \sum_\kb \alpha^{ij}_\kb \right|^2 = \underbrace{\sum_\kb \left|\alpha^{ij}_\kb\right|^2}_\text{direct terms} + \underbrace{\sum_{\substack{\kb,\kb' \\ \kb\neq\kb^\prime}} \left(\alpha^{ij}_\kb\right)^* \alpha^{ij}_{\kb^\prime}.}_\text{interference terms}
\end{split}\label{eq:direct_interference}
\end{align}
The interference terms can be constructive or destructive.
If enough electronic transitions with a finite amplitude are in phase we detect a large Raman intensity.
However, if the contributions are out of phase, interference will lead to a small or even zero Raman intensity.
The weight of the direct terms in the final result is much smaller than that of the interference terms (see Supporting Information).

The key point of this paper is to use the concept of quantum interference to explain the observed behavior of the Raman intensity with laser energy in \mote.
This concept was shown to be important in the Raman intensities of graphene where an increase of the Raman intensity is observed when destructive interference terms are Pauli-blocked through electron or hole doping.\cite{basko_calculation_2009,kalbac_influence_2010,chen_controlling_2011,reichardt_ab_2017}
We show that selection rules manifest themselves at the level of quantum interference, but even when selection rules do not apply, quantum interference explains the behavior of the
Raman intensity.
Since interference effects reflect the interplay of all the terms $\alpha^{ij}_\kb(\omega)$, it is inaccurate to attribute the features in the behavior of the Raman intensities to a single electronic transition. 

\section{Results}\label{sec:results}

\subsection{Experimental results}

\begin{figure*}[!tbh]
\begin{center}
\includegraphics[width=0.8\linewidth]{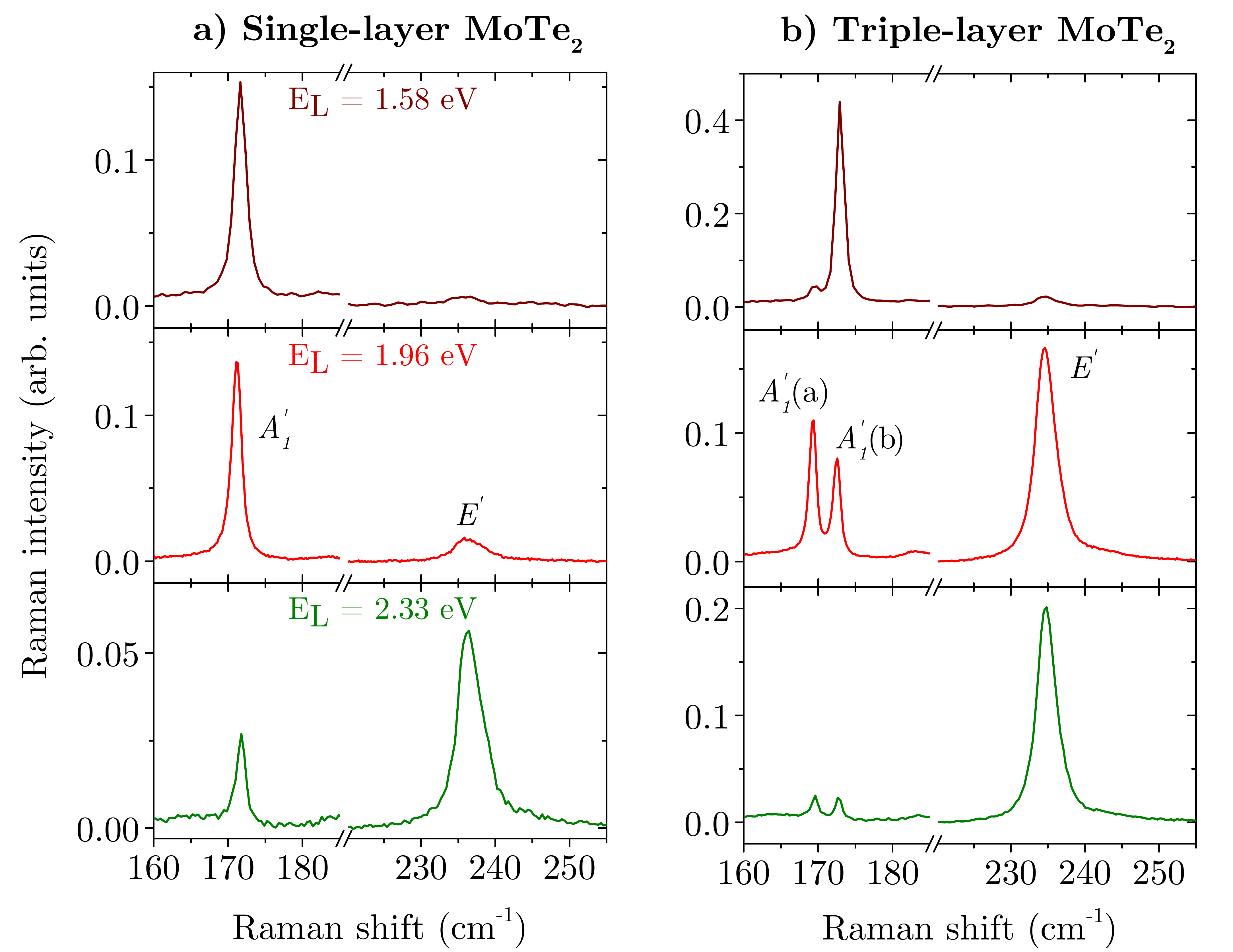}
\caption{Micro-Raman spectra of single-layer (panel a) and triple-layer (panel b) \mote at three different laser energies in a backscattering geometry. All the spectra have been normalized by the integrated intensity of the Raman mode from the underlying Si substrate at $\approx 520~\rm cm^{-1}$. The corresponding atomic displacements for the Raman-active modes are shown as insets in the upper panels.}
\label{fig:Fig_IPCMS}
\end{center}
\end{figure*}

Single- and few-layer hexagonal \mote (hereafter simply denoted \mote) samples were prepared by mechanical exfoliation and deposited onto Si substrates covered with a 90 nm SiO$_2$ epilayer.
The Raman spectra of single- and triple-layer \mote were measured at three different laser energies ($E_{\rm L}=1.58~\rm eV$, 1.96~eV, and 2.33~eV) in a backscattering geometry using a custom-built micro-Raman setup.
The incoming laser beam was linearly polarized and the Raman scattered light was sent through a monochromator with a 500 mm focal length coupled to a charge-coupled device (CCD) array.
A 900 (resp. 2400) lines/mm grating was used for measurements at 1.58 eV (resp. 1.96 eV and 2.33 eV).
Laser intensities below $50~\rm kW/cm^{-2}$ were employed in order to avoid photoinduced heating and sample deterioration.
The Raman spectra were fit with Voigt profiles taking into account the spectral resolution of our setup of 1.0, 0.4 and  0.6~cm$^{-1}$ at $E_{\rm L}=$1.58~eV, 1.96~eV, and 2.33~eV, respectively.
Figure~\ref{fig:Fig_IPCMS} shows micro-Raman spectra of single- and triple-layer \mote.
The number of \mote layers has been unambiguously identified as described in Ref.\citenum{froehlicher2015}.
The raw spectra have been normalized by the integrated intensity of the T$_{\rm 2g}$ (point group O$_{\rm h}$) Raman mode of silicon at $\approx 520~\rm cm^{-1}$ for a qualitative comparison.
To quantitatively compare experimentally measured Raman intensities with the \abinitio Raman susceptibilities calculated according to Eq.~\ref{eqI}, we have also taken optical interference effects into account and extracted the $\rm xx$-component of the Raman susceptibility after carefully considering the polarization-dependent response of our setup. Additional details on the normalization procedure can be found in the Supporting Information.


In Figure.~\ref{fig:Fig_IPCMS}, we show the experimentally obtained Raman spectra of single- (panel (a)) and triple-layer (panel (b)) MoTe$_2$. The prominent A$^\prime_1$ and E$^\prime$~modes are clearly visible. In single-layer MoTe$_2$, the A$^\prime_1$~mode dominates the spectrum at laser energies of $E_L$=1.58 and 1.96~eV, while at $E_L$=2.33~eV the E$^\prime$~mode is dominant.
Similarly, in triple-layer \mote, the \apr and \epr mode dominate the Raman spectra at $E_{\rm L}=1.58~\rm eV$ and $E_{\rm L}=2.33~\rm eV$, respectively. However, the  \apr modes feature and the \epr mode have comparable intensities at $E_{\rm L}=1.96~\rm eV$.

Remarkably, the Davydov-split \apa and \apc modes have similar intensities in triple-layer \mote at $E_{\rm L}=1.96~\rm eV$ and $2.33~\rm eV$, whereas the bulk-like \apc mode is 13 times more intense than the \apa mode at $E_{\rm L}=1.58~\rm eV$.  Note that the \epr mode does not display a measurable Davydov spitting.\cite{froehlicher2015} In Figure~\ref{fig:raman}, we will compare the experimentally measured Raman susceptibilities and integrated intensity ratios between the \apc and \apa modes with \abinitio calculations and correlate the observation of a prominent Davydov splitting with resonantly enhanced Raman intensities.

\subsection{Theoretical calculations}

In the following we will discuss the results of first-principles calculations and compare them with our experimental results.
Before discussing the results for triple-layer \mote, we analyze the single-layer case.
This allows us to introduce the concept of quantum interference in a simpler context.
In all cases we will analyze the $\rm xx$-component of the Raman susceptibility tensor, $\alpha^\mu_{\rm xx}(\omega)$.
The other components are related to $\rm xx$-component of $\alpha$, as shown in Table~\ref{tab:phonon_modes}.

\subsubsection{Single-layer \mote}

\begin{figure}[h!]
\center
\includegraphics[width=0.45\textwidth]{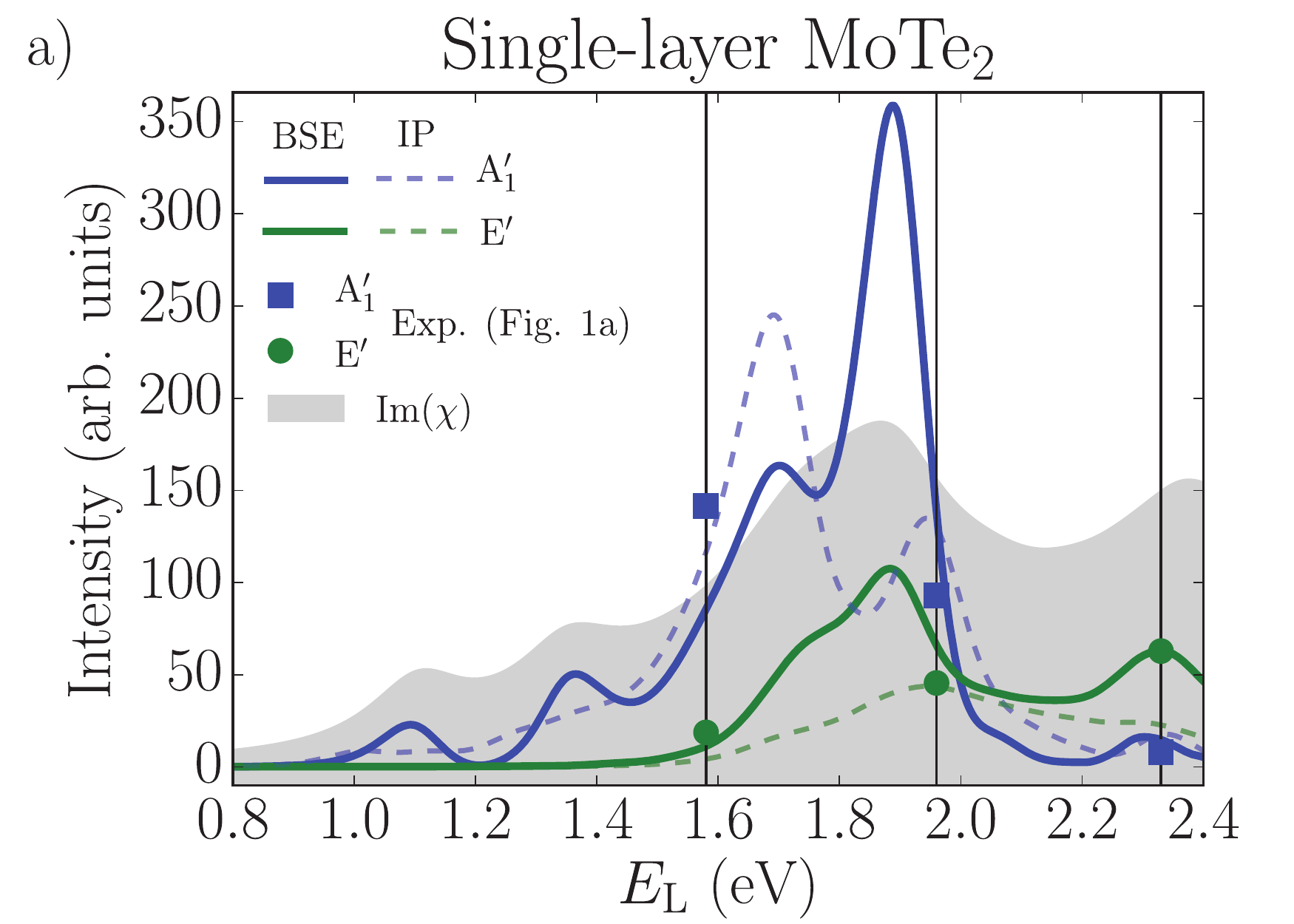}
\includegraphics[width=0.45\textwidth]{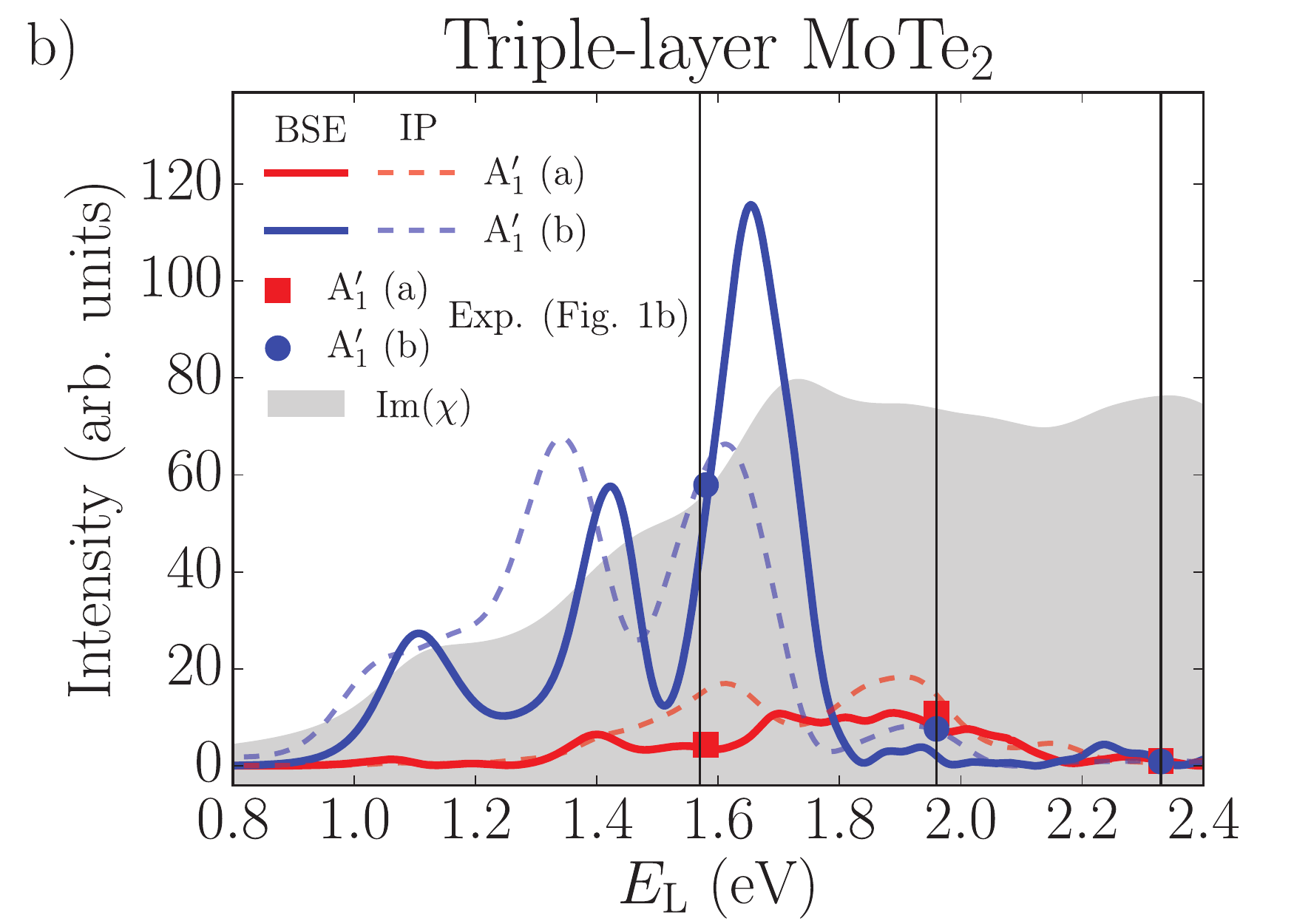}
\includegraphics[width=0.45\textwidth]{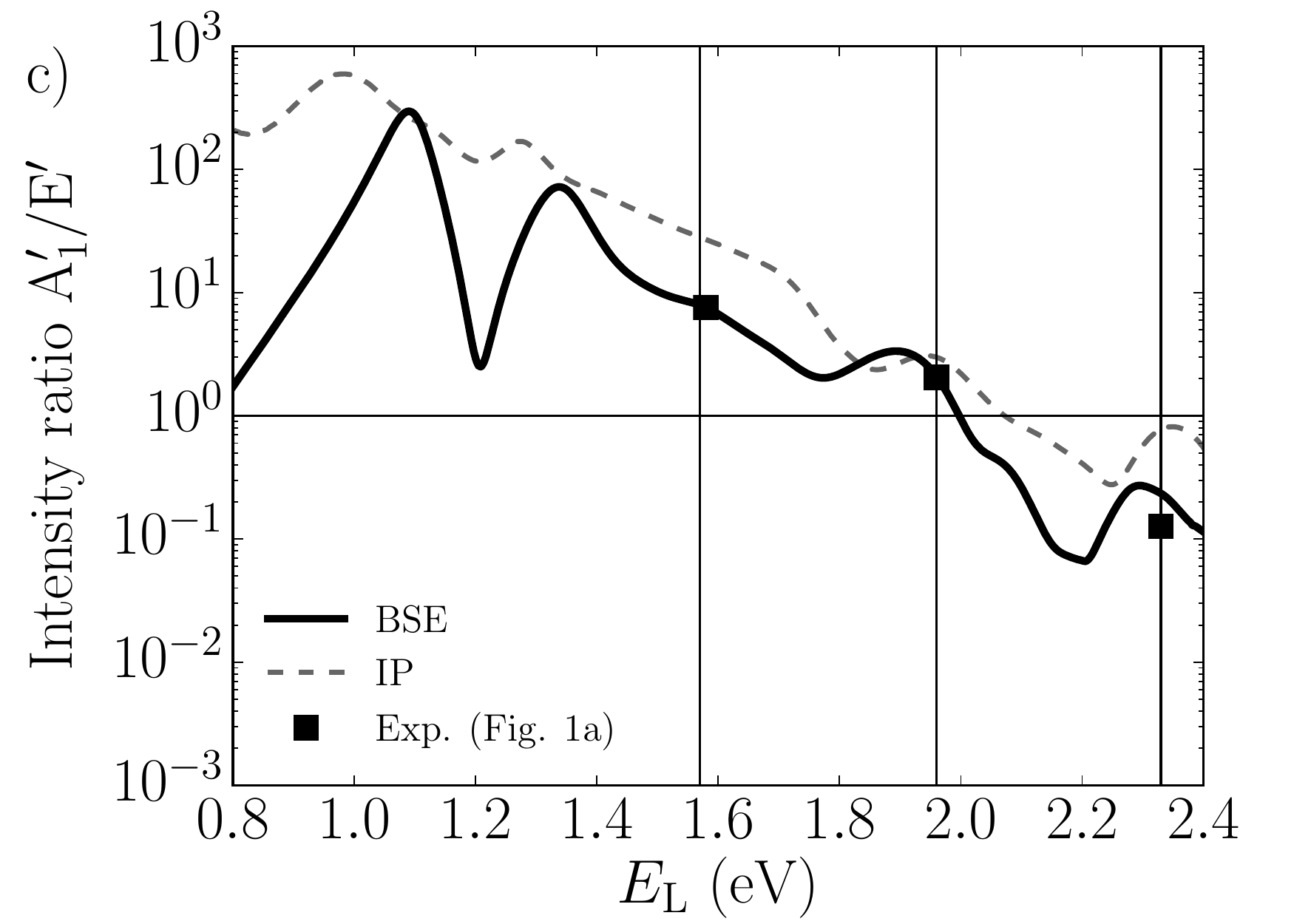}
\includegraphics[width=0.45\textwidth]{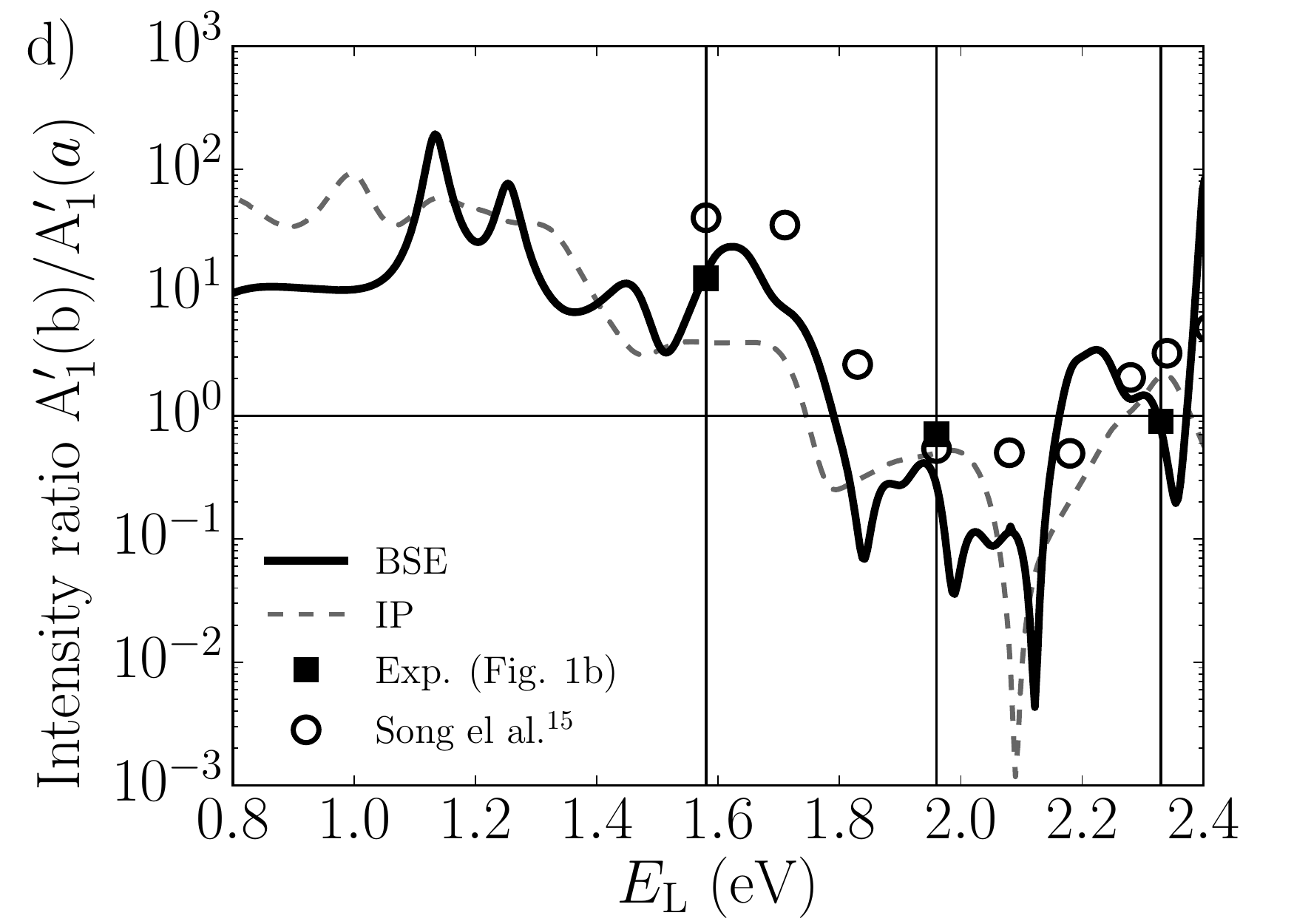}
\caption{
a) and b) Calculated $\rm xx$-component of the Raman susceptibility tensor squared ($|\alpha^{\rm xx}|^2$) at the IP~level (dashed line) and at the BSE~level (solid lines) for single-layer (panel a) and triple-layer (panel c) \mote as a function of laser energy for the $A^\prime_1$(a) and $A^\prime_1$(b) modes.
The blue squares and green circles correspond to the same quantity (up to a normalization factor) extracted from the spectra in Figure~\ref{fig:Fig_IPCMS}a and b using Eq.~{\ref{eqI}}.
The vertical lines are guides to the eye.
The BSE optical absorption is represented by a gray area. The optical gap is in good agreement with the experimental values reported in Refs.~\citenum{ruppert_optical_2014} and \citenum{froehlicher_direct_2016}.
c) and d) Ratio of the intensities of the $A^\prime_1$ and $E^\prime$ modes (panel c) and $A^\prime_1$(b) and $A^\prime_1$(a) modes (panel d) calculated on the IP~level (dashed line) and BSE~level (solid line).
The black squares represent to the experimentally observed ratios.
}
\label{fig:raman}
\end{figure}

In the case of single-layer \mote, we analyze the Raman susceptibility for the $A^\prime_1$ and $E^\prime$ modes.
Figure~\ref{fig:raman}a shows the Raman susceptibility as a function of laser energy for both the IP (dashed lines) and BSE calculations (solid lines).
Up to a laser energy of 2~eV the intensity of the $A^\prime_1$ mode is larger than that of the $E^\prime$ mode.
At higher laser energy, the $E^\prime$ mode has a larger intensity than the $A^\prime_1$ mode, in good agreement with the experimental data reported here and in the literature.\cite{ruppert_optical_2014,froehlicher2015,grzeszczyk_raman_2016}
The overall scale of the theoretical results (IP- and BSE-level) is has been chosen to reflect that of the experimental results.
This chosen scale is the same for both the IP- and BSE-level calculation to allow a comparison between the two.
Since the overall scaling factor cancels when considering intensity ratios, the quantity that can be compared unambiguously between experiment and theory is the ratio of two intensities, as shown in Figure \ref{fig:raman}c and d.
The inclusion of many-body effects does not change this general trend but affects the relative intensities at the excitonic transitions.

We first analyze the contributions of the individual $\kb$-points to the IP susceptibility $\chi_\kb(\omega)$.
Figure \ref{fig:1l_raman_bands}a shows $\chi_\kb(\omega)$ along a path through the high-symmetry points in the BZ.
The main contributions to $\chi(\omega)$ for laser energies between 0.8 and 2~eV come from the lower bands in transition space in a region around K and between K and M.

\begin{figure}[h!]
\includegraphics[width=0.5\textwidth]{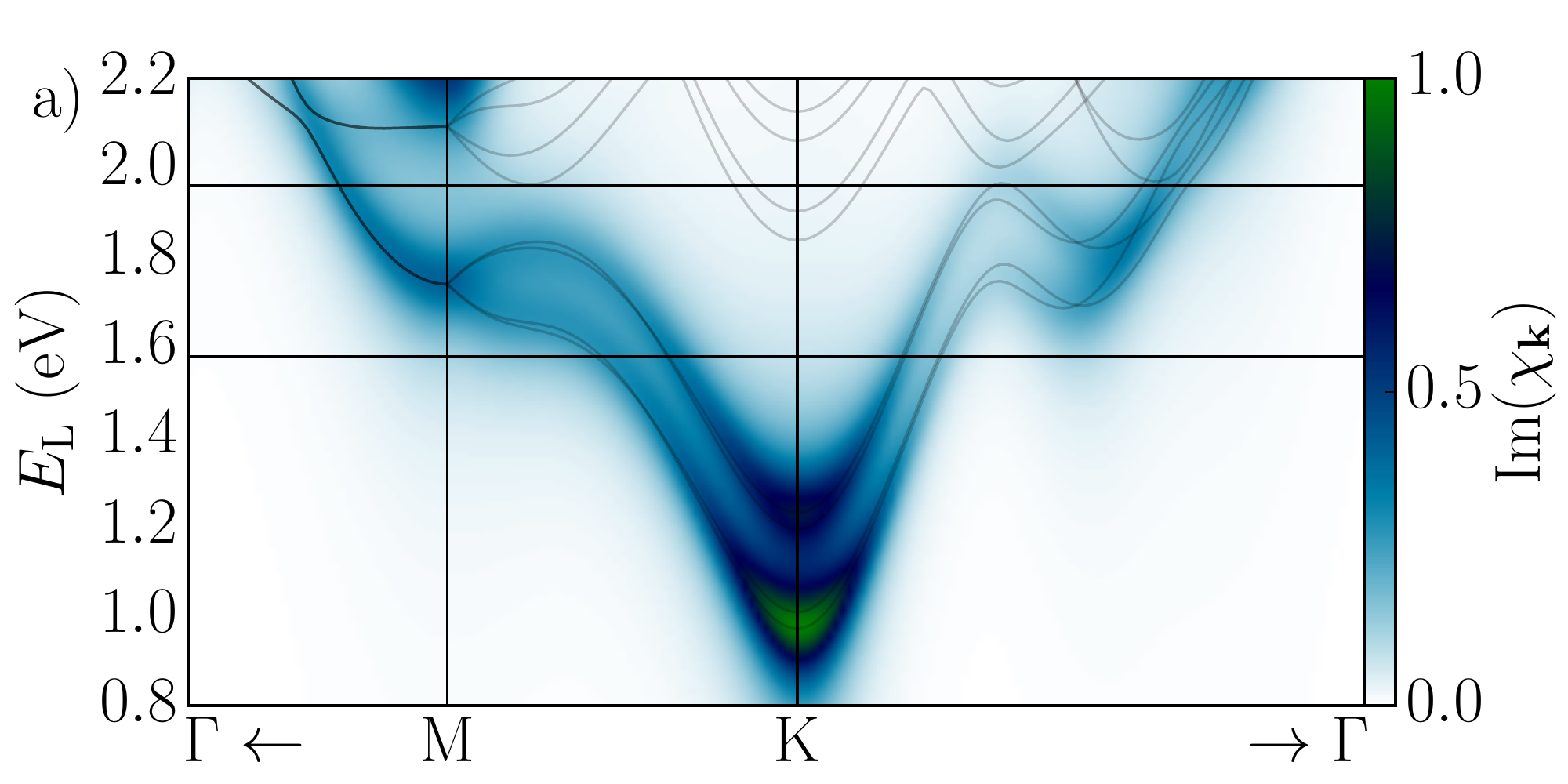}
\includegraphics[width=0.5\textwidth]{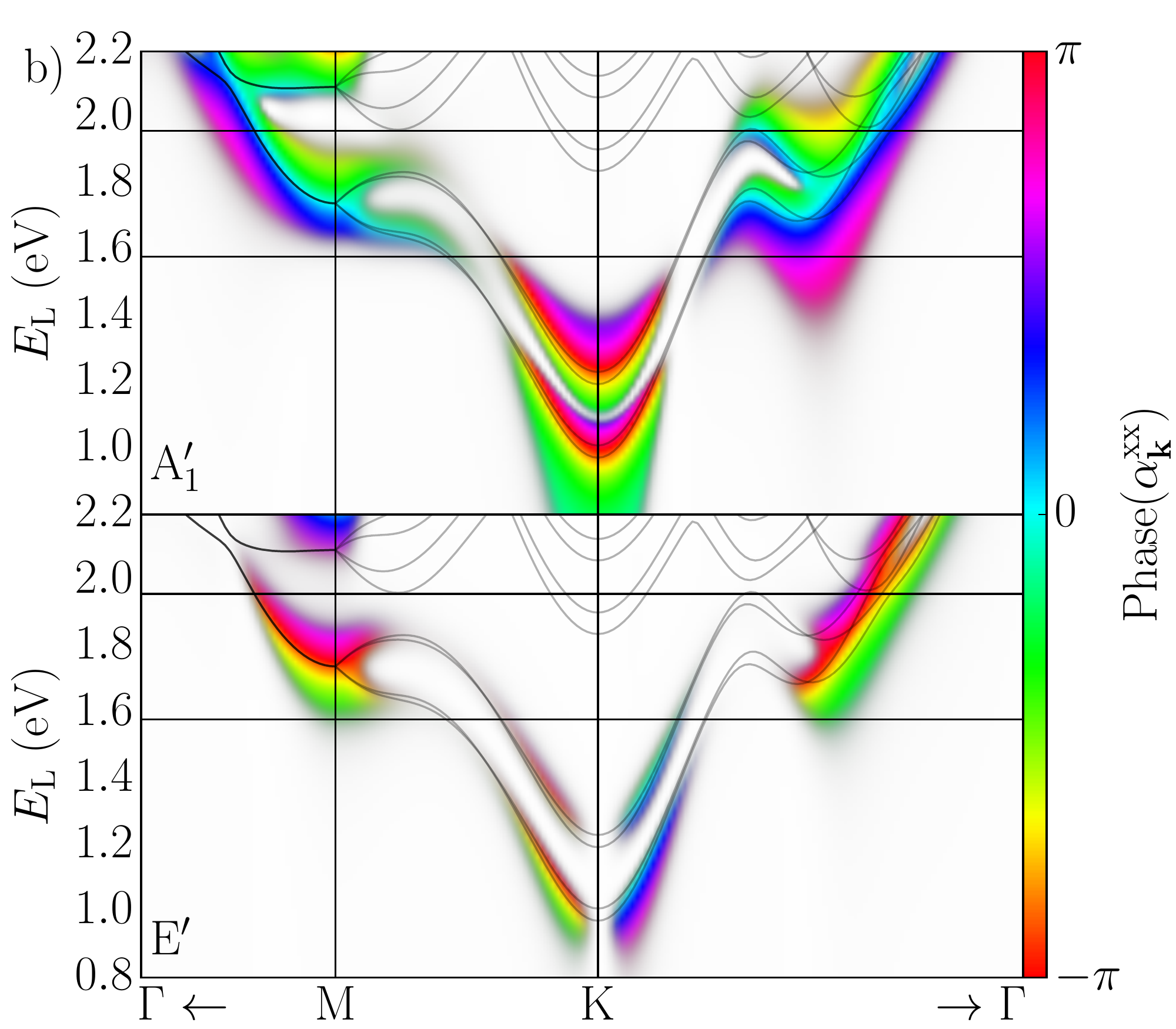}
\includegraphics[width=0.5\textwidth]{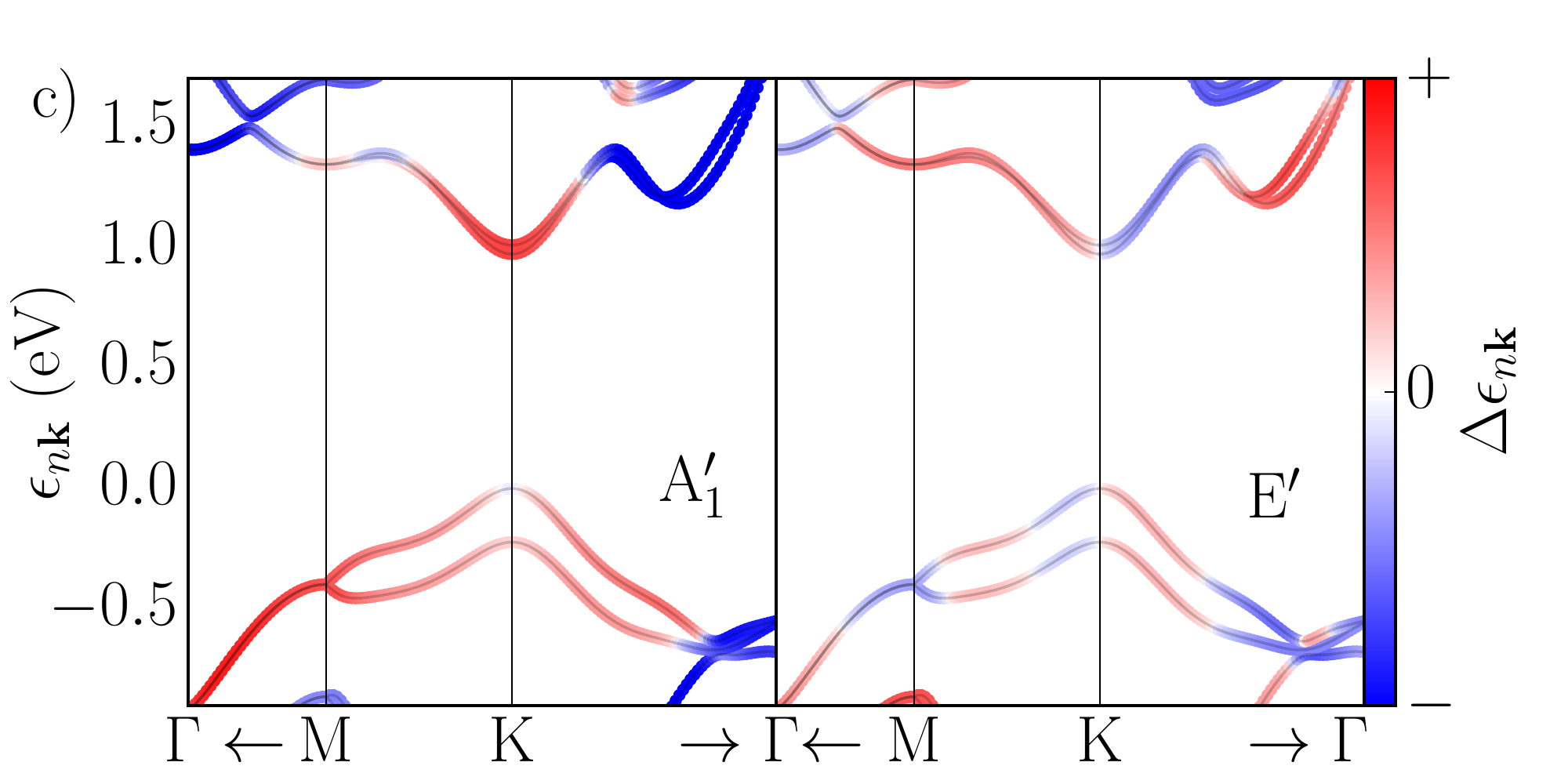}
\caption{
a) IP absorption Im($\chi^{\rm xx}_\kb$) represented in transition space along the high-symmetry points in the Brillouin zone for single-layer \mote.
We only show points close to M and K as there are no relevant transitions close to the $\Gamma$ point for laser energies up to 2~eV. 
b) Raman susceptibility $\alpha^{\rm xx}_\kb$($\omega$) along the high-symmetry line.
Points at which the absolute value of $\alpha^{\rm xx}_\kb$($\omega$) is below 7\% of the maximum value at that $\omega$ are shown in white, otherwise the phase of $\alpha^{\rm xx}_\kb$($\omega$) is represented by color.
The horizontal lines correspond to the laser energies used in our experiment.
c) Change of electronic bands with atomic displacements according to the $A^\prime_1$ and $E^\prime$ phonon modes.}
\label{fig:1l_raman_bands}
\end{figure}

It should be noted that only optically active transitions can contribute to the Raman susceptibility, but not all of them necessarily do so.
For instance, near the band gap, the $A^\prime_1$~mode is active while the $E^\prime$~mode is silent, even though the same electronic transitions contribute and both modes are, in principle, allowed by lattice symmetry.
This behavior can be understood in terms of angular momentum conservation.
Near the band gap at K the band structure is rotationally symmetric and thus angular momentum is conserved.
Both incoming and outgoing photons carry an angular momentum of $\pm\hbar$ while the $E^\prime$ phonon does as well.
This implies that the final state has a total angular momentum of $\pm2\hbar$ or $0$, which violates angular momentum conservation and renders the $E^\prime$~mode silent.
By contrast, the phonon corresponding to the $A_1^\prime$~mode does not carry angular momentum and hence the corresponding process is allowed.

This can also be understood from the point of view of quantum interference.
For this purpose, we show the $\kb$-resolved Raman susceptibly as a function of $\omega$ in Figure~\ref{fig:1l_raman_bands}b.
We color-encode the phase when the amplitude is larger than 7\% of the maximum amplitude at that laser energy.
For the $E^\prime$ mode, the positive contribution from one side of the valley is added to the negative contribution from the other side,
which leads to an overall cancelation of the Raman intensity.
By contrast, for the $A^\prime$ mode the contributions add up constructively.
At higher laser energies, the full rotation symmetry gradually gets broken down to the 120$^\circ$ rotation symmetry of the lattice, an effect known as trigonal warping\cite{saito_trigonal_2000} of the electronic structure.
Angular momentum is then only conserved up to integer multiples of $\pm3\hbar$ and both the \apr and \epr modes become allowed.

In order to track down the origin of the phase of the Raman susceptibility, we take a closer look at the derivative of $\chi_\kb$($\omega$) with respect to atomic displacements:\cite{cardona_light_scattering_solids_II}
\begin{align}
\frac{\partial\chi^{ij}_\kb(\omega)}{\partial Q} \propto& \Bigg\{
\frac{\partial(\Delta \epsilon_{cv\kb})}{\partial Q} \frac{(\Lambda^i_{cv\kb})^* (\Lambda^j_{cv\kb})}
{(\omega - \Delta \epsilon_{cv\kb} + i\gamma)^2}
+ \frac{\partial \left[(\Lambda^i_{cv\kb})^*(\Lambda^j_{cv\kb}) \right]}{\partial Q} \frac{1}
{\omega - \Delta \epsilon_{cv\kb} + i\gamma} 
+ (\omega \to -\omega)\Bigg\}.
\end{align}
where $\Delta \epsilon_{cv\kb} = \epsilon_{c\kb} - \epsilon_{v\kb}$. 

The first term involves the change of the electronic band energies, which is given by the diagonal (intra-band) electron-phonon coupling (EPC) matrix elements.
The second term stems from the change of the ELC upon atomic displacements and involves the off-diagonal EPC matrix elements.
The first term is double-resonant and corresponds to a process where an electron is excited to a conduction band, then scatters with a phonon within the same band, and finally decays to the valence band by emitting a photon.
Since this term is double-resonant, we assume it to be dominant and we can directly relate the phase of the Raman susceptibility with the sign of the diagonal EPC matrix elements.
We visualize these by plotting the change of the electronic band energies with respect to atomic displacements, which correspond to the diagonal EPC matrix elements, as shown in Figure~\ref{fig:1l_raman_bands}c.
From this plot we observe a direct correlation between the sign of the diagonal EPC and the phase of the Raman susceptibility in Figure~\ref{fig:1l_raman_bands}b.
Therefore, we attribute constructive or destructive interference between regions of the BZ to differences in sign of the change of the electronic band energies.

\subsubsection{Triple-layer \mote}

In the case of triple-layer \mote, we focus our attention on the \apa and (b) modes, for which experiments reported here and in the literature\cite{froehlicher2015,grzeszczyk_raman_2016,song_physical_2016} show a variation of the relative Raman intensity as a function of laser energy.
Our calculations, both on the IP and BSE level, reproduce this observation very well, as shown in Figure~\ref{fig:raman}b and d.
Common to both calculations is that the \apa phonon mode is dominant in intensity for laser energies up to 1.8 eV while at higher laser energies the \apc mode is dominant.
However, only with the inclusion of excitonic effects (BSE) do we obtain the experimentally observed ratio.
Contrary to the single-layer case, where the different intensities are related to different symmetries of the phonon modes, in the triple-layer case, the \apa and (b) modes belong to the same representation and hence symmetry based-arguments do not apply.
However, we can still use the concept of quantum interference introduced previously to explain the intensity inversion.

\begin{figure}
\center
\includegraphics[width=0.5\textwidth]{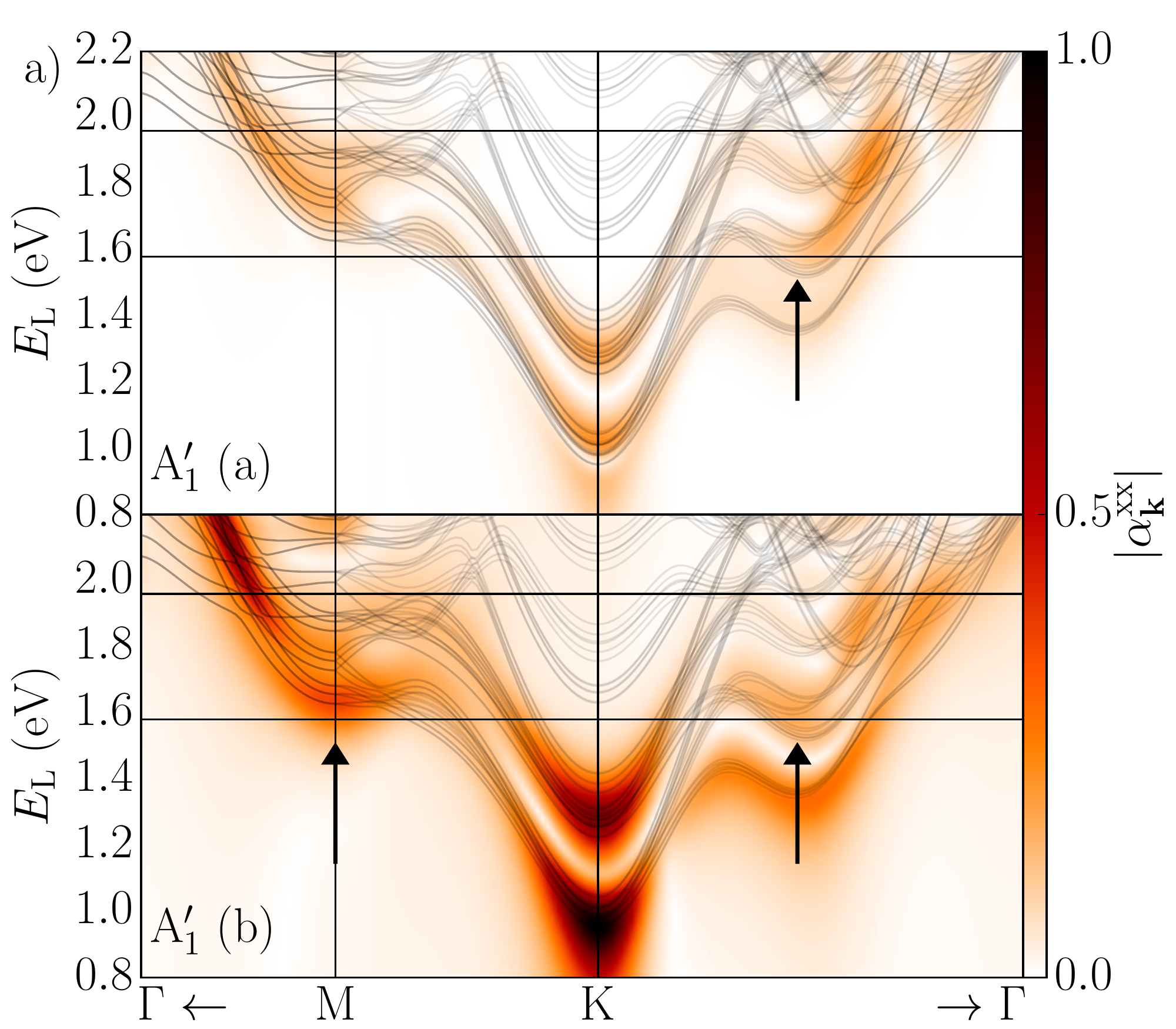}
\includegraphics[width=0.5\textwidth]{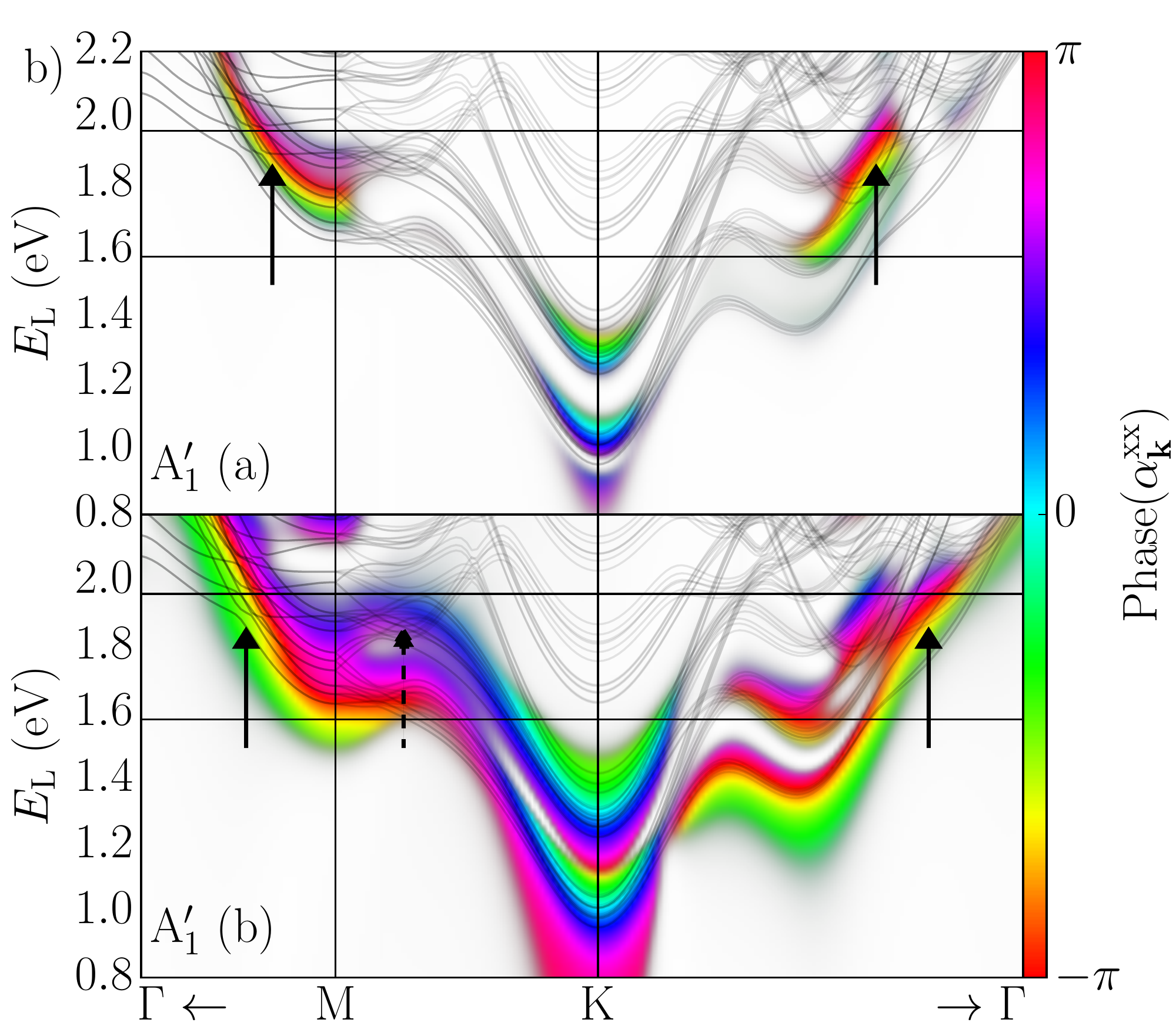}
\includegraphics[width=0.5\textwidth]{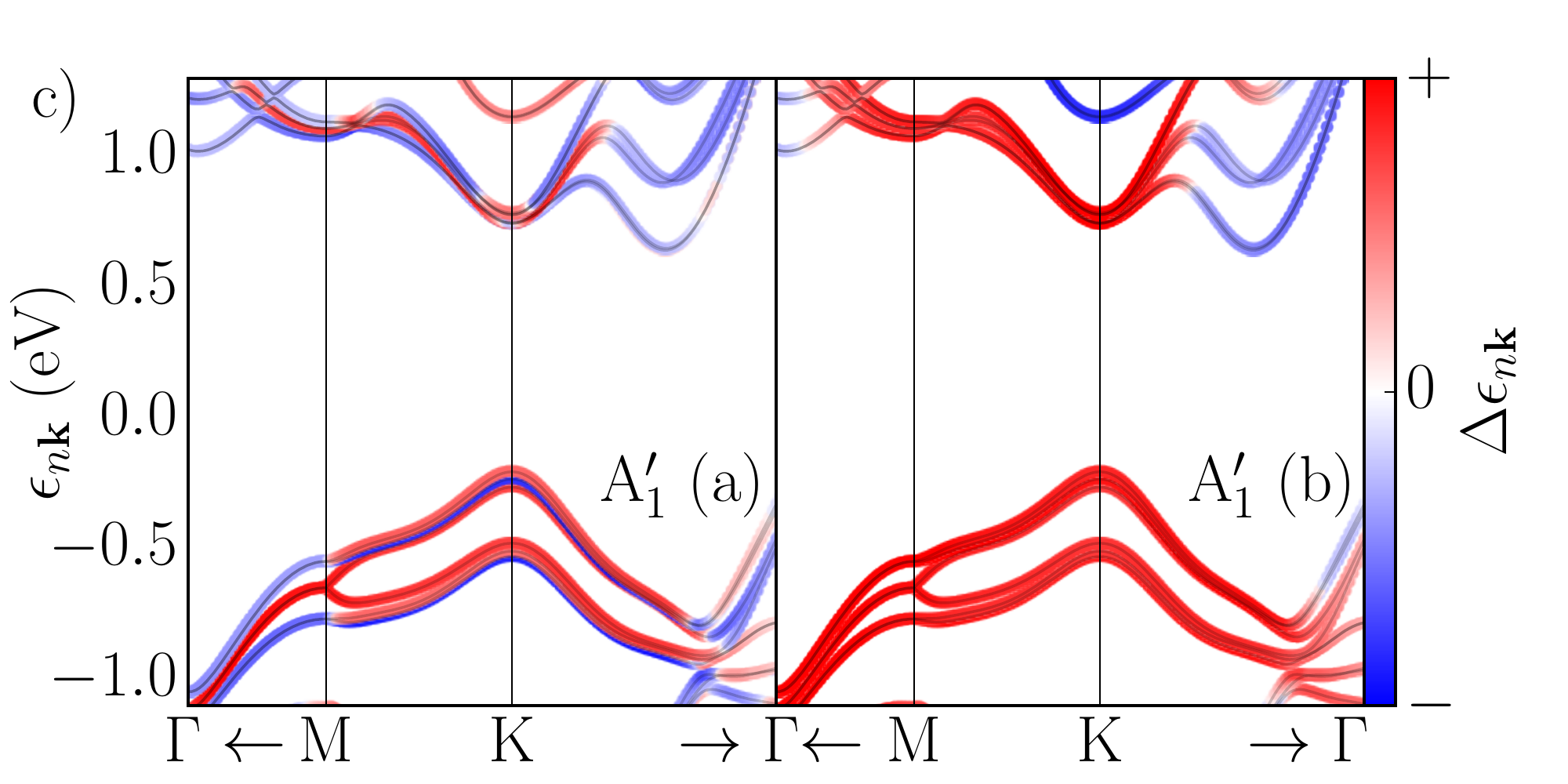}
\caption{
$\kb$-point resolved contributions $\alpha^{\rm xx}_\kb$($\omega$) to the total Raman susceptibility for triple-layer \mote.
Panel (a) shows the absolute value, while panel (b) shows the phase of $\alpha^{\rm xx}_{\kb}$($\omega$).
The phase is only shown if the absolute value if greater than 7\% of the maximum value at that $\omega$.
Panel (c) shows the change of the electronic bands with atomic displacements according to the \apa (left) and \apc~modes (right).}
\label{fig:3l_raman_bands}
\end{figure}

We start by analyzing the behavior of the Raman susceptibility for laser energies near the band gap energy.
There, the $A_1^\prime$(b)~mode has a large intensity while the \apa~mode is practically silent.
This can be understood from Figure~\ref{fig:3l_raman_bands}c, where we show the diagonal EPC matrix elements along the high-symmetry line in the BZ.
For the  $A_1^\prime$(b)~mode the conduction band states at K contribute with the same sign, while for the \apa~mode they have opposite signs.
This is a direct consequence of the band composition at K and the way the layers vibrate (in-phase in the $A^\prime_1$(b) mode and out-of-phase in the $A^\prime_1$(a) mode, respectively - see Supporting Information)
.

\begin{figure}[h!]
\center
\includegraphics[width=0.4\textwidth]{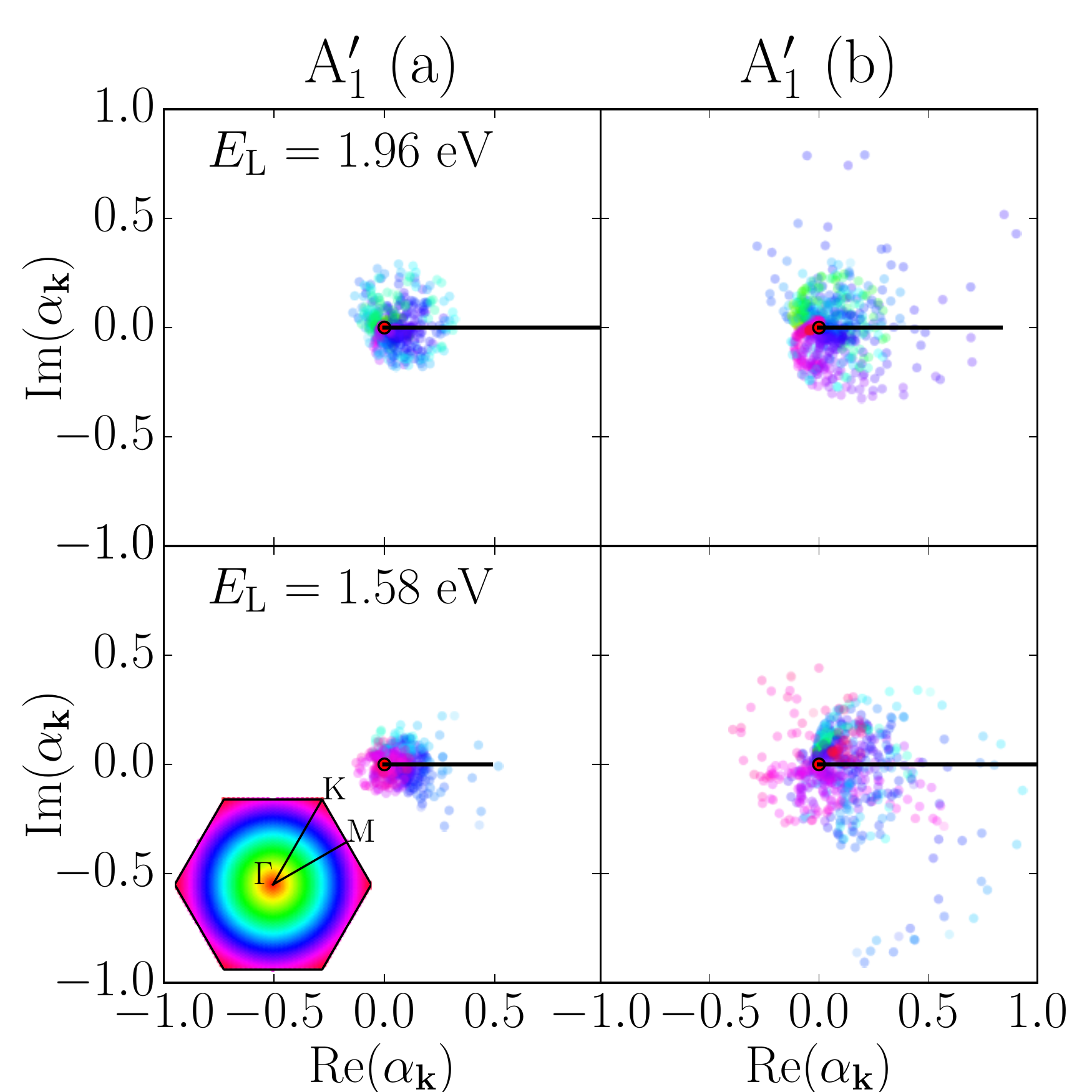}
\caption{
Argand plot of $\alpha_\kb$($\omega$) for the $A_1^\prime$(a) and (b)~modes of triple-layer \mote for laser energies $E_{\rm L}$=1.58~eV (bottom panel) and 1.96~eV (top panel).
The colors represent the position of the point in the Brillouin zone (see inset).}
\label{fig:3l_interference}
\end{figure}

The Raman intensities at higher laser energies (between 1.58 and 1.96 eV) can also be understood from the point of view of quantum interference.
For this, we represent the contributions from all $\kb$-points in the BZ as points in the complex plane (``Argand plot'') as shown in Figure~\ref{fig:3l_interference}.
By color-encoding the $\kb$-point location in the BZ, we can identify the regions which contribute constructively to the total Raman amplitude and those that are interfering destructively.
The overall phase of the different contributions has been fixed such that the total Raman susceptibility is real and positive (solid black line).
At a laser energy of 1.58~eV, the contributions from the edge of the BZ, i.e., between K and M (purple dots), scatter concentrically around the origin and mostly cancel each other for both phonon modes.
However, the regions between K and $\Gamma$ and M and $\Gamma$ (blue dots) are building the signal up.
Since these contributions have larger amplitude for the $A_1^\prime$(b)~mode than for the \apa mode, the former has to a larger intensity at this laser energy. 
This becomes clear by looking at Figure~\ref{fig:3l_raman_bands}, where we represent the absolute value of $\alpha_\kb(\omega)$ along the high-symmetry line in panel (a) and its phase in panel (b).
For a laser energy of 1.58~eV, there are resonant transitions between K and $\Gamma$ and at M (see arrows in panel (a)).
At these points the modulus of $\alpha_\kb(\omega)$ is large and the phases are the same, which leads to constructive interference of the signal and an increase in the observed Raman intensity for both phonon modes.

At a laser energy of 1.96~eV, the situation is rather different.
The $\alpha_\kb(\omega)$ contributions from the region between K and M (purple dots in the Argand plot) no longer scatter concentrically around the origin and now destructively interfere with the contributions from the K-$\Gamma$ and M-$\Gamma$ regions (blue dots).
We resolve which electronic transitions lead to these destructive interference effects by referring once more to Figure~\ref{fig:3l_raman_bands}.
The destructive contributions stem from transitions at M, which have a relative phase of $\pi/3$ (blue areas in Figure~\ref{fig:3l_raman_bands}b) while the constructive ones have relative phases between $-\pi/2$ and $-\pi$ (green, yellow, and red areas).

In the case of the~\apa mode, the amplitude of these destructive contributions is small and hence the resulting signal is larger than the one of the $A_1^\prime$(b)~mode, for which the destructive contributions have a sizable amplitude.
From Figure~\ref{fig:3l_raman_bands}a we can verify that both the amplitude of the $\alpha_\kb(\omega)$ near the M point is larger for the $A_1^\prime$(b)~mode and that their phases are opposite to the ones from the contributions of the constructively interfering points (see dashed and solid arrows in panel (b)).

The reason for the small amplitudes in the K-M region for the \apa~mode can be deduced from Figure~\ref{fig:3l_raman_bands}c.
The diagonal EPC matrix elements for the \apa~mode and the lowest conduction bands along the K-M direction have both positive and negative signs.
Consequently, their contribution to $\alpha_{\kb}$($\omega$) mostly cancels out, which leads to a small contribution to the Raman susceptibility.
On the other hand, for the $A_1^\prime$(b)~mode, the different EPC matrix elements add up with the same sign and the $\kb$-points from this region give a larger contribution. 

\section{Conclusions and Outlook}

We calculated the laser energy-dependent Raman susceptibility in an \textit{ab initio} framework by taking finite differences of the dynamic dielectric susceptibility in the frozen-phonon approximation.
We applied our method to study the Raman spectrum of single- and triple-layer \mote, reproducing and explaining the experimentally observed behavior of the intensity ratio as a function of laser energy for the different $A^\prime_{1}$ phonon modes.
We demonstrated that quantum interference effects between contributions of electronic transitions from different parts of the Brillouin zone are responsible for this behavior.
We also found a correlation between the phase of these contributions and the sign of the diagonal electron-phonon coupling matrix elements.
Quantum interference effects make the direct correlation of the optical absorption spectrum with the measured and calculated Raman intensities highly non-trivial.
Additionally, we showed that symmetry arguments are not always enough to explain the counterintuitive behavior of the intensities as a function of laser photon energy as seen in the case of the $A^\prime_1$ modes of triple-layer MoTe$_2$.
Instead, a careful and detailed analysis is required to trace down which features of the electronic structure, vibrational spectra, and interplay between them are responsible for the observed behavior.
Furthermore, we showed that the proper inclusion of excitonic effects is necessary to accurately describe the experimentally observed intensity ratio of the modes as a function of laser energy.
The approach presented here offers a way to systematically analyze resonant Raman spectra.
Because of its \textit{ab initio} nature, it can be directly used to study different phonon modes of various materials in different phases.
Additionally, it can also be applied to study the temperature dependence of the Raman spectrum, as recently investigated experimentally.\cite{golasa_resonant_2017} This could be done by including the electron lifetimes and renormalization from electron-phonon coupling as recently shown for the temperature dependent optical absorption of MoS$_2$.\cite{molina-sanchez_temperature-dependent_2016}

\begin{acknowledgement}
We thank Etienne Lorchat for fruitful discussions. 
The authors acknowledge support by the National Research Fund, Luxembourg (Projects OTPMD, RAMGRASEA, C14/MS/773152/FAST-2DMAT, and INTER/ANR/13/20/NANOTMD) and the Agence Nationale de la Recherche, France (under grant H2DH ANR-15-CE24-0016). S.B. is a member of the Institut Universitaire de France (IUF).
The simulations were done using the HPC facilities of the University of Luxembourg~\cite{VBCG_HPCS14}.
The authors declare no competing financial interests.
\end{acknowledgement}

%
%
\newpage
\begin{center}
{\Huge\textbf{Supporting Information}}
\end{center}

\setcounter{equation}{0}
\setcounter{figure}{0}
\setcounter{section}{0}
\renewcommand{\theequation}{S\arabic{equation}}
\renewcommand{\thefigure}{S\arabic{figure}}
\renewcommand{\thesection}{S\arabic{section}}

\section{Ground state properties and phonons}

Calculations of the electronic ground-state properties were done within density functional theory (DFT) in the local density approximation (LDA). Since LDA is known to underestimate the lattice parameters, we use the experimentally determined lattice constant of MoTe$_2$, $a$=3.52~\AA\cite{podberezskaya_crystal}. We chose an LDA exchange-correlation function over more elaborate van~der~Waals functionals, as it has been shown to perform well in predicting vibrational properties of layered materials.\cite{luo_effects_2013,luo_anomalous_2013}

\begin{figure}
	\centering
  \includegraphics[width=.45\textwidth]{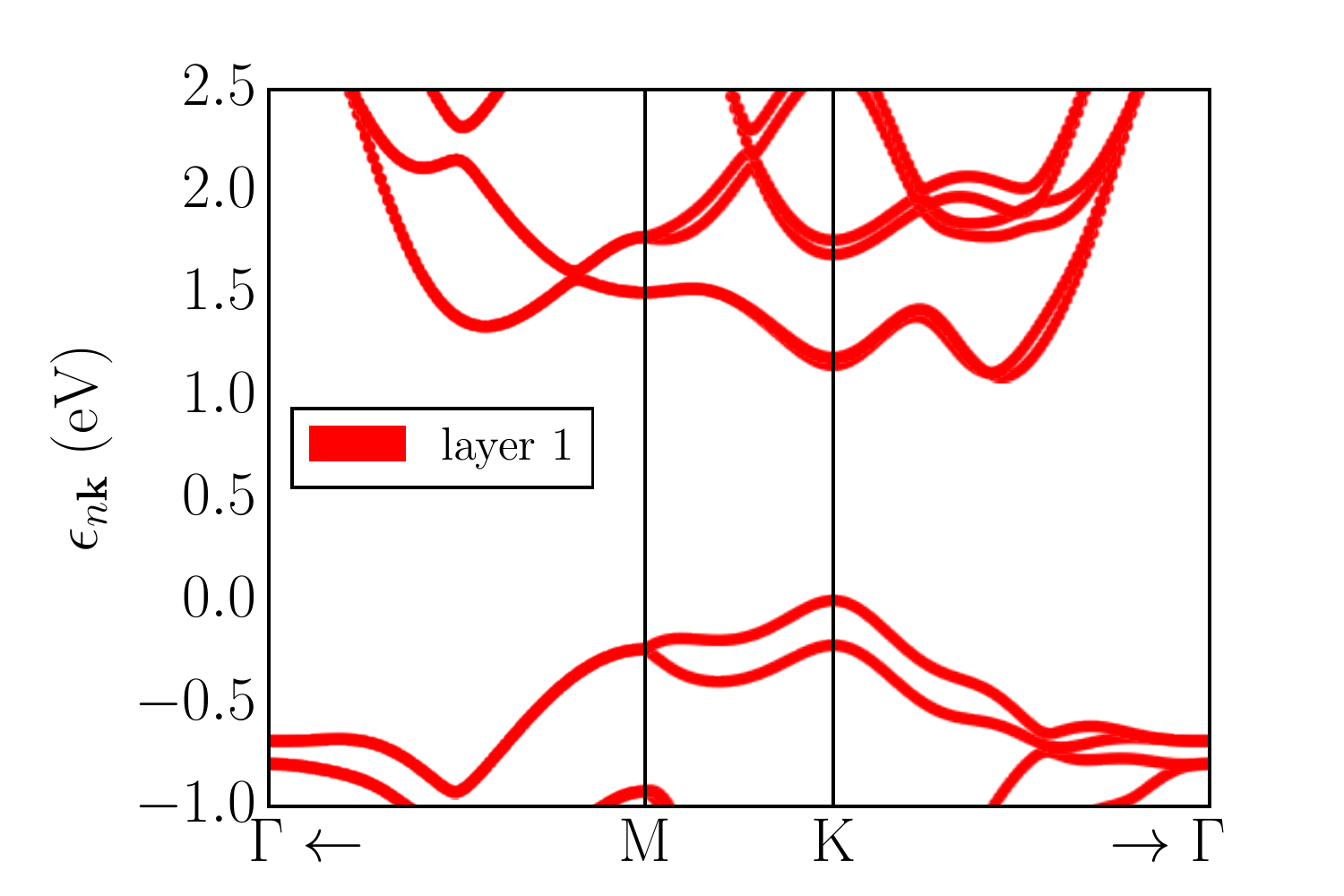}
  \includegraphics[width=.45\textwidth]{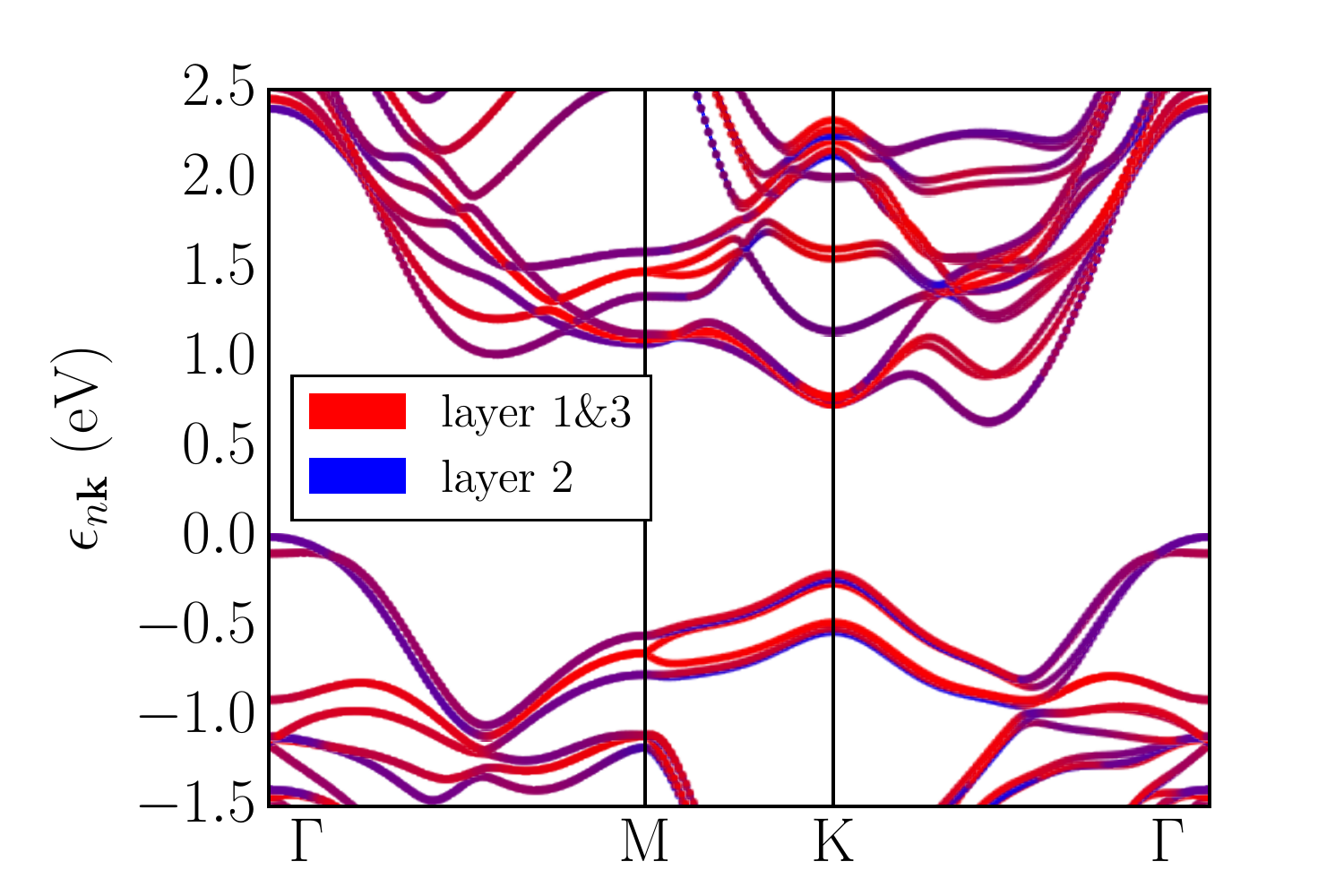}
\caption{Band structures of single- and triple-layer \mote in the LDA approximation including spin-orbit coupling.}\label{fig:bands}
\end{figure}

Figure \ref{fig:bands} shows the electronic band structure of single- (left panel) and triple-layer (right panel) \mote.
When passing from the single- to the triple layer case, each single-layer band splits into a triplet of bands. We represent the contributions of the different layers to the orbital composition of each band by color (red for the outer layers 1 and 3 and blue for the inner layer 2).
This decomposition should be compared to Figure~4c in the main text, where the sign of the band energy change with atomic displacements according to the A$^\prime_1$(a) and (b) phonon modes is shown.
In the case of the A$^\prime_1$(b)~mode, the three layers vibrate in phase (see inset in Figure~1b of the main text) and hence the band energies within each band triplet always change with the same sign, independent of the layer composition of the bands.
For the A$^\prime_1$(a)~mode, on the other hand, the oscillation phase of the inner layer is opposite to that of the two outer layers (see inset in Figure~1b of the main text) and due to the different layer contributions to each band triplet member, the sign of the band energy changes varies within the triplet.

\section{Optical absorption}

We calculated the GW quasi-particle correction to the LDA eigenvalues using the
\texttt{yambo} code.\cite{marini_yambo:_2009} We used a $36\times36\times1$ sampling of the Brillouin
zone (BZ) for single- and triple-layer \mote. We used a 40 Ry cutoff for the
plane-wave basis set, a Coulomb cutoff technique\cite{rozzi_exact_2006} to avoid spurious interactions
between the periodic copies in the z-direction and a vacuum separation of 50 and
70 Bohr  for single- and triple-layer, respectively.

We calculated the GW quasiparticle corrections for the band gap and applied a scissor shift\cite{gonze_dynamical_1997} to the LDA band energies of the other bands to account for this corrections without having to compute them explicitly.
The scissor operator is kept fixed for the different atomic displacements.
This approximation has the advantage that only one calculation of the correction of the band gap 
energy is needed. However, it does not account for the changes of the screening effects in the electron-phonon 
interaction.
A consistent way of including these corrections is still desirable and will be the topic of future work.

\begin{table}
	\begin{tabular}{cc}
	        & scissor shift (eV)\\\hline
	    Single-layer    & 0.667\\
	    Triple-layer    & 0.548\\
	\end{tabular}
	\caption{Scissor operator for single- and triple-layer \mote.}
	\label{tab:scissor}
\end{table}

The calculation of the dielectric susceptibility including many-body effects has
been performed by solving the Bethe-Salpeter equation (BSE) with the \texttt{yambo} code.\cite{marini_yambo:_2009}
The static dielectric screening was calculated using the same vacuum separation between the layers as in the GW case.
The number of electronic transitions included to construct the BSE Hamiltonian was selected to include electronic transitions inside a energy window of 3~eV.
We find this criterion to be more meaningful physically and the convergence of the spectra to be more stable compared to selecting the number of valence and conduction bands separately.
Especially in the triple-layer case, we find many dispersive and crossing bands near the lower conduction and topmost valence band (see Figure~\ref{fig:bands}b), which makes it difficult to know \textit{a priori} how many bands need to be included in the calculation.
Additionally, checking the convergence with the gradual inclusion of valence and conduction bands can lead to a false convergence of the dielectric susceptibility.

\section{Results}

\subsection{Single-layer}

As complementary information to the main text, we represent the intensity of the individual contributions $\alpha_\kb(\omega)$ in transition space for the two phonon modes (A$_1^\prime$ and E$^\prime$) of single-layer MoTe$_2$ in Figure~\ref{fig:1l_raman_bands}.

\begin{figure}[h!]
\includegraphics[width=0.5\textwidth]{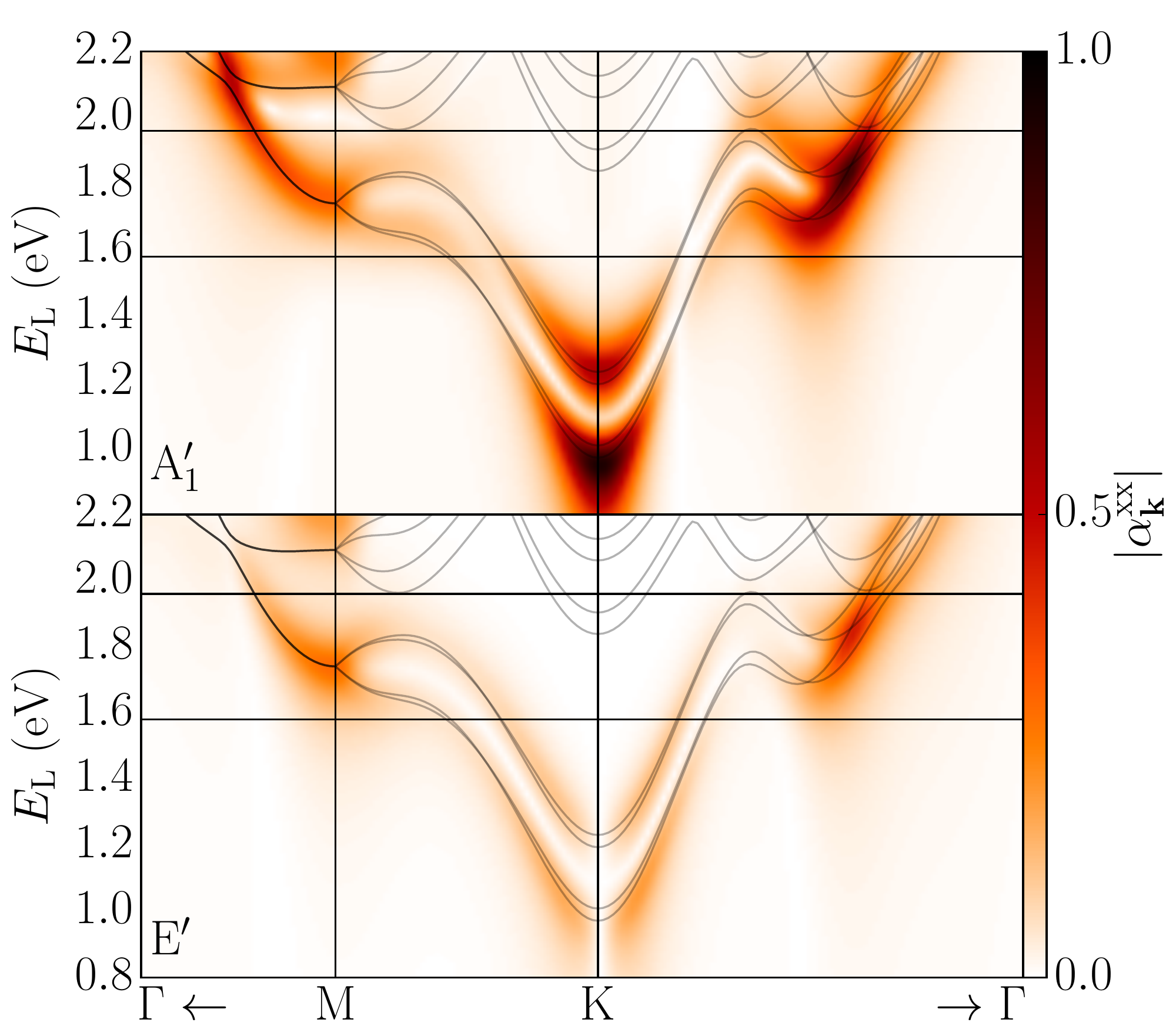}
\caption{Absolute value of the Raman susceptibility resolved along the high-symmetry $\kb$-points $\alpha^{\mathrm{xx}}_\kb$($\omega$) for single-layer \mote.}
\label{fig:1l_raman_bands}
\end{figure}

\begin{figure}[h!]
\center
\includegraphics[width=0.8\textwidth]{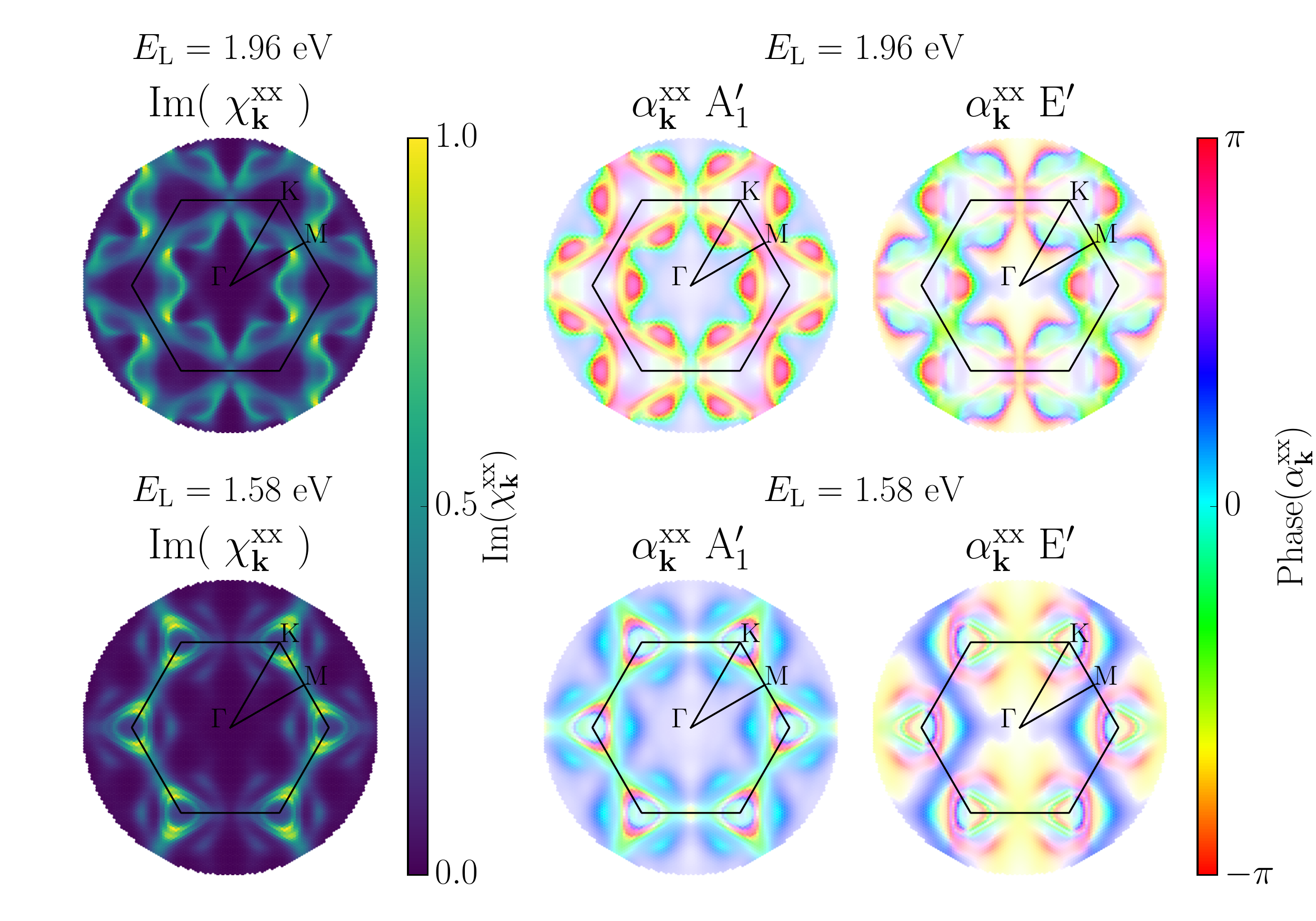}
\caption{
$\kb$-point-resolved contributions to the absorption spectrum $\chi^{\rm xx}_\kb$ (left panel) and Raman susceptibility $\alpha^{\mathrm{xx}}_\kb$($\omega$) (right panel) for single-layer MoTe$_2$ across the BZ for two different laser energies used in experiment. The E$^\prime$~mode was chosen to be polarized in the y-direction (compare Raman tensors in Table~1 of the main text).
}
\label{fig:1l_raman_bz}
\end{figure}

It is also instructive to look at the individual contributions $\alpha_\kb(\omega)$ over the full Brillouin zone (FBZ) as shown in Figure \ref{fig:1l_raman_bz}.
A line cut along the high-symmetry points of this is shown in Figure~3 of the main text.
However, there are additional contributions from regions not along the high-symmetry, shown in Figure \ref{fig:1l_raman_bz}.
In all cases, we consider the contributions for incoming and outgoing light polarized along the x-direction.
This leads to a breaking of some symmetries of the lattice and to the emergence of two inequivalent M and K points.
We choose to represent the contributions along the high-symmetry line represented in Figure~\ref{fig:1l_raman_bz} to 
simplify the analysis without compromising the main conclusions.
In the case of the $E^\prime$~mode the phonon was chosen to be polarized along the x-direction.
 
\subsection{Triple-layer}

\begin{figure}
\center
\includegraphics[width=0.5\textwidth]{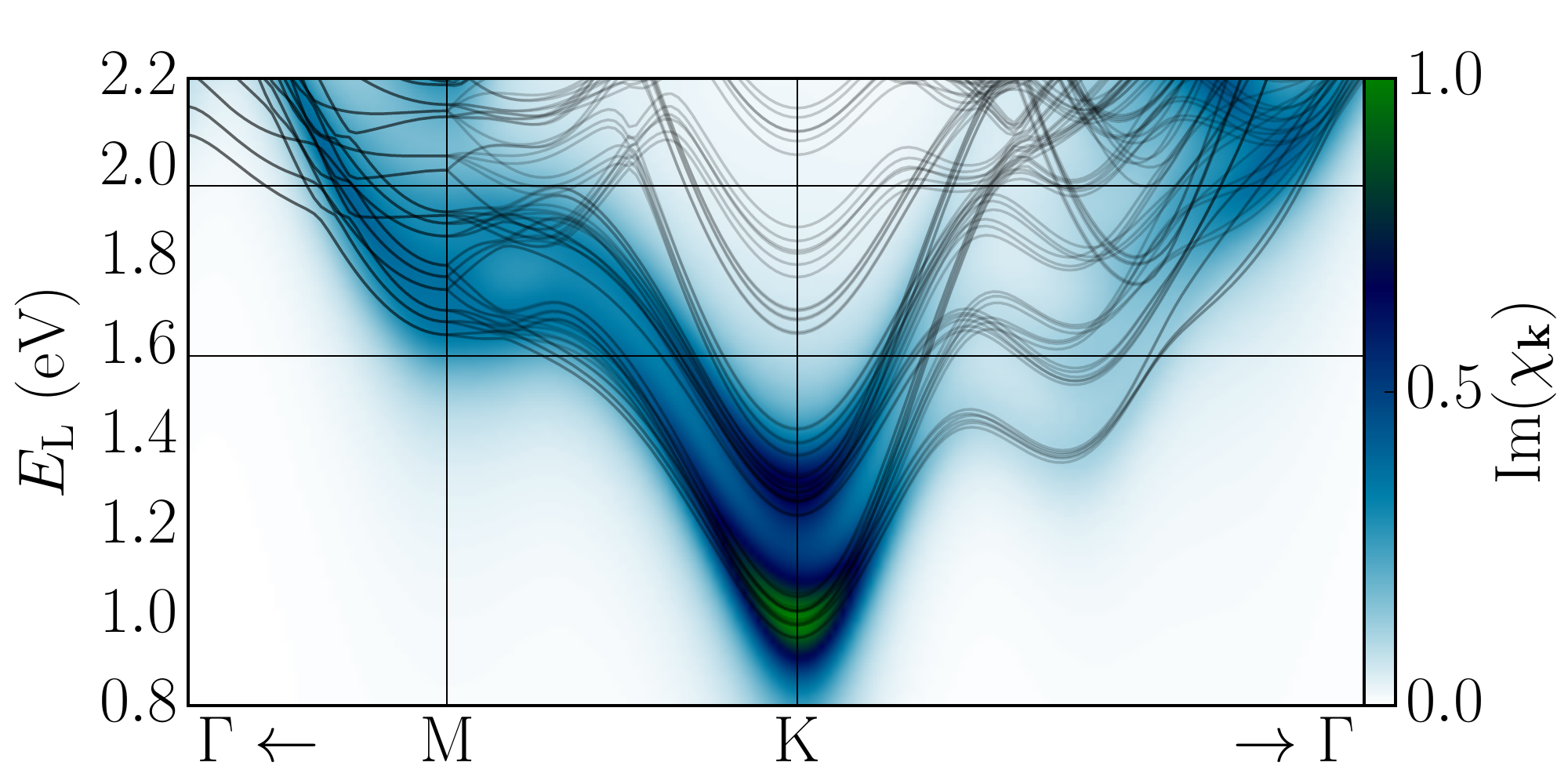}
\caption{
IP absorption Im\{$\chi_\kb(\omega)\}$ represented in transition space along the high-symmetry points in the Brillouin zone for triple-layer \mote.
}
\label{fig:3l_chi}
\end{figure}

\begin{figure}
\center
\includegraphics[width=0.8\textwidth]{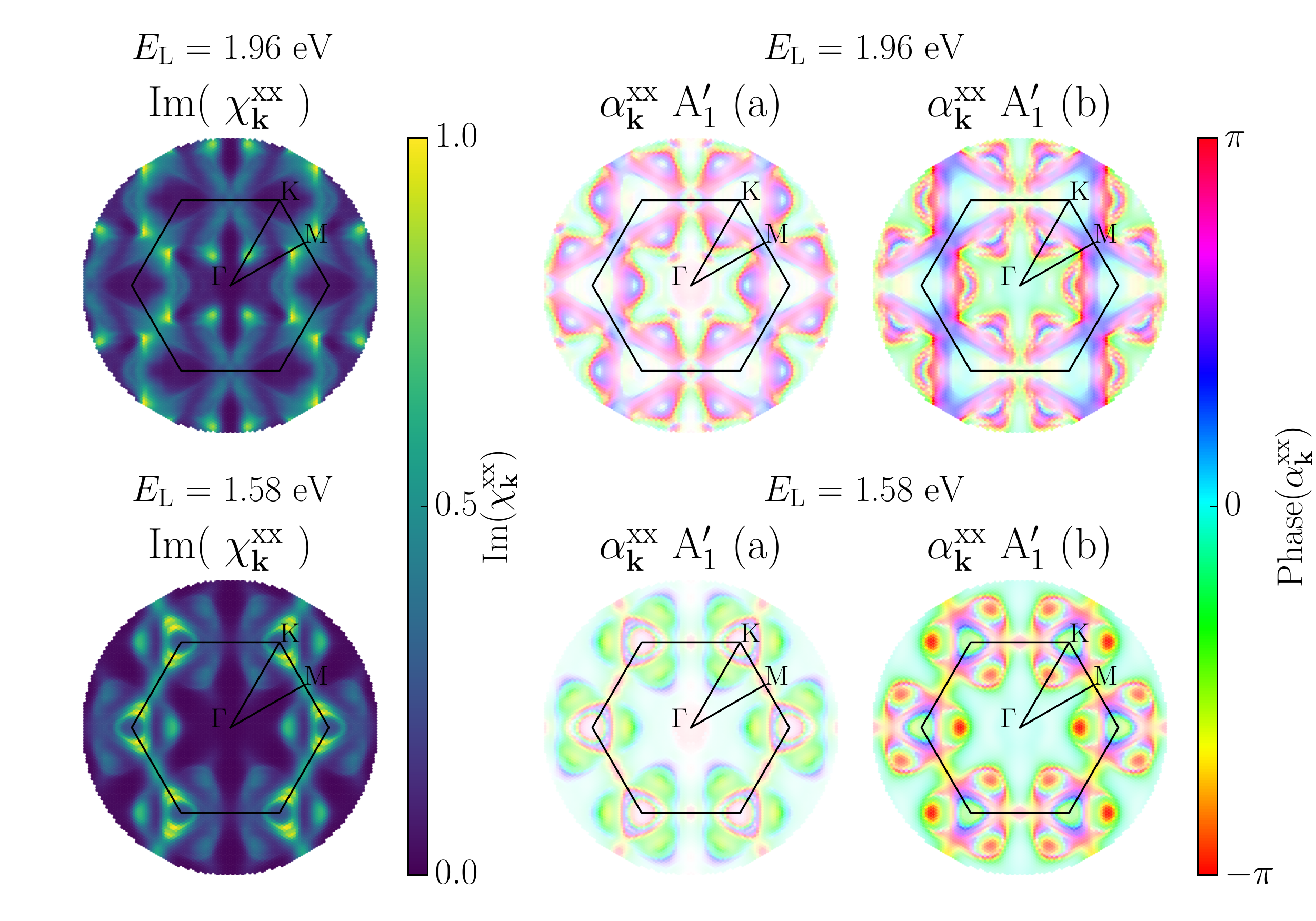}
\caption{
$\kb$-point-resolved contributions to the absorption spectrum $\chi^{\rm xx}_\kb$ (left panel) and Raman susceptibility $\alpha^{\mathrm{xx}}_\kb$($\omega$) (right panel) for single-layer MoTe$_2$ across the BZ for two different laser energies used in experiment.
} \label{fig:3l_raman_bz}
\end{figure}

For the triple-layer case, we represent Im$\{\chi^{\rm xx}_\kb(\omega)\}$ along the high-symmetry line in Figure~\ref{fig:3l_chi}.
We additionally represent the individual contributions $\alpha_\kb(\omega)$ on the full BZ as represented in Figure~\ref{fig:3l_raman_bz} for the two energies (1.57~eV and 1.96~eV) used in our experiments.
A line cut through this figure along the high-symmetry points is shown in Figure~4a of the main text.
Similarly to the $A_1^\prime$ mode in single-layer, the symmetry is broken along the x-direction.

\section{Direct and interference terms}

We performed calculations with and without the interference terms in the IP level.
The omission of the interference terms leads to Raman intensities that are orders of magnitude smaller than those obtained by including them.
This is consistent with the fact that the calculation of the interference terms involves two integrations over the Brillouin zone compared to only one integration for the ``direct'' terms (see Equation 8 in the main text). Thus their weight compared to the ``direct'' terms is in general much larger.
Ignoring the interference terms leads to the absence of the observed intensity inversion of the Davydov multiplet of the A$_1^\prime$ modes.

\begin{figure}
\center
\includegraphics[width=0.6\textwidth]{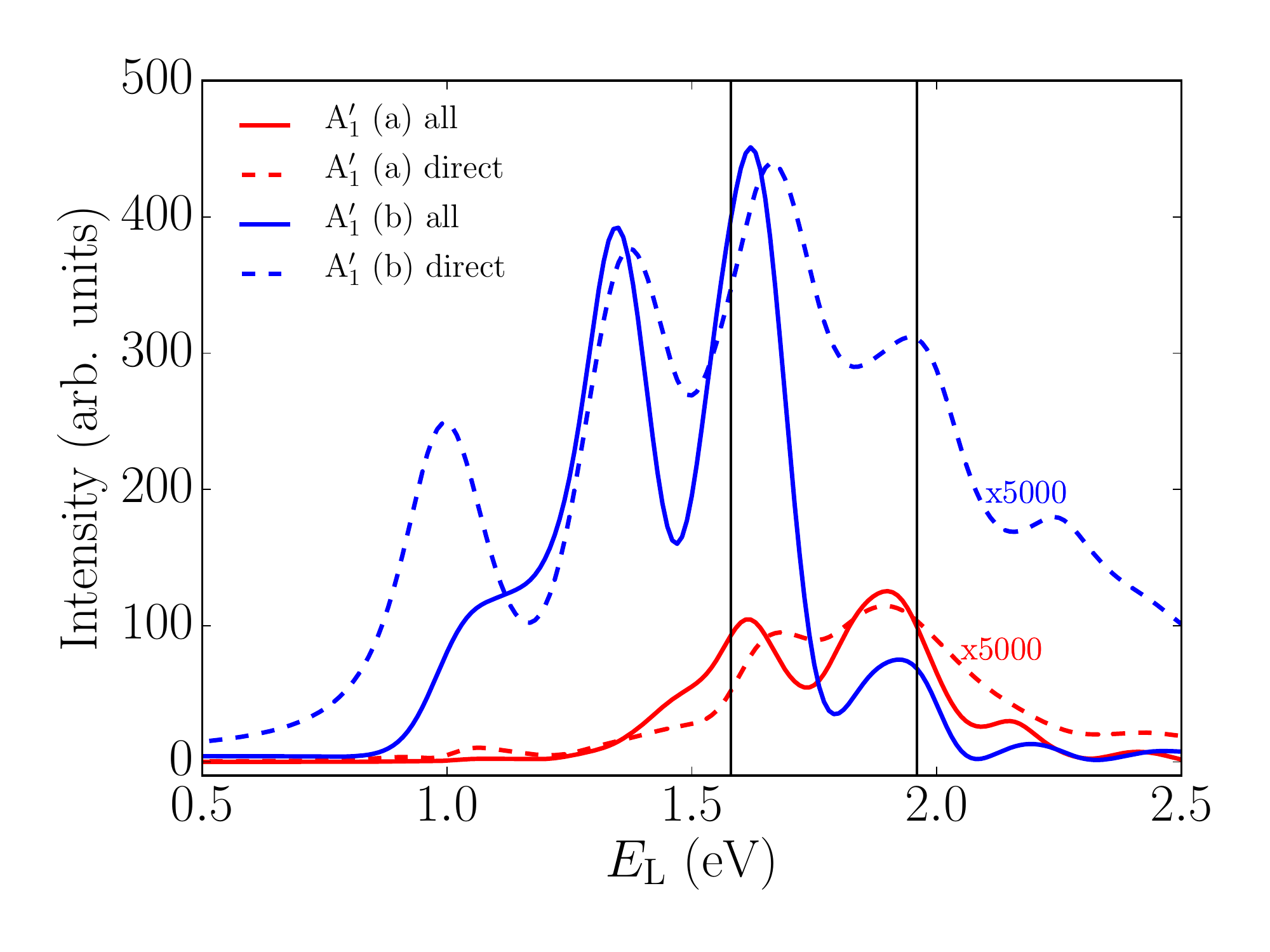}
\caption{ 
Relative contributions of the ``direct'' (dashed line) and ``interference+direct'' (solid line) terms to the total Raman susceptibility for triple-layer \mote. The distinction between direct and interference terms is explained in Equation 8 in the main text.
} \label{fig:3l_raman_interference}
\end{figure}


\section{Normalization procedure for the experimental data}

To quantitatively compare Raman susceptibilities recorded at different laser photon energies, one has to carefully normalize the Raman spectra. Indeed, the spectra may not be acquired under the exact same conditions (e.g., different integration time, laser intensity,\dots) and the detection efficiency of the experimental setup may also be different. To get rid of all these dependencies, one can normalize the measured Raman intensities to the integrated intensity of a close-lying and well-known Raman feature. We chose the Raman mode of the bulk silicon substrate at around $\sim 520~{\rm cm}^{-1}$, which has been very well documented, for instance in Ref.\cite{Renucci1975}. Furthermore, one also has to take into account the dependence of the measured Raman intensity on the laser photon energy as well as optical interference effects.

Consequently, the normalized Raman intensity of a given Raman mode X is given by
\begin{equation}
\left.\frac{I_{\rm X}}{I_{\rm Si}}\right|_{\rm normalized}(E_{\rm L})=\left(\frac{E_{\rm Si}}{E_{\rm X}}\right)^3 \frac{F_{\rm Si}(E_{\rm L},E_{\rm Si})}{F_{\rm X}(E_{\rm L},E_{\rm X})}C_{\rm Si}(E_{\rm L})\left.\frac{I_{\rm X}}{I_{\rm Si}}\right|_{\rm measured}(E_{\rm L},E_{\rm X},E_{\rm Si}), \label{eq_norm_X_Si}
\end{equation}
where $E_{\rm L}$ is the incoming laser photon energy, $C_{\rm Si}$ is a coefficient that takes into account the resonance effect in the Si mode intensity as shown in Ref.\cite{Renucci1975}, $I_{\rm X}$ and $I_{\rm Si}$ are the integrated intensity of the X and Si mode, $E_{\rm X}$ and $E_{\rm Si}$ are the energies of the Raman scattered photons contributing to the X and Si modes, and $F_{\rm X}$ and $F_{\rm Si}$ are the enhancement factors for the X and Si modes in the [Si/SiO$_2$/single- or triple-layer MoTe$_2$/air] layered system, respectively. Note that after applying Eq.~\eqref{eq_norm_X_Si}, the integrated intensity ratio $\left.\frac{I_{\rm X}}{I_{\rm Si}}\right|_{\rm normalized}$ only depends on $E_{\rm L}$. Let us also note that the $\left(\frac{E_{\rm Si}}{E_{\rm X}}\right)^3$ term stems from the photon energy dependence of the Raman scattered energy flux ($\propto E^4$) and from the fact that our detector -a charge-coupled device (CCD) array- measures a signal proportional the number of incoming photons, not to the energy flux.
In the range of energies studied here, the coefficient $C_{\rm Si}$ is directly deduced from Figure~6 in Ref.\cite{Renucci1975}. The enhancement factors are obtained following Yoon \textit{et al.}\cite{Yoon2009} and Soubelet \textit{et al.}\cite{soubelet_resonance_2016}. To obtain reliable enhancement factors, we have first carefully estimated the refractive index of few-layer \mote from the measurement of the intensity of the Si Raman mode in a [Si/SiO$_2$/$N$-layer \mote/air] layered system as a function of the number of layers $N$, similarly to Zhang \textit{et al.}\cite{Zhang2015e}. Second, to accurately estimate the measured Raman signal from the Si substrate, we have considered the semi-transparency of bulk Si and the fact that we use a confocal Raman setup. Indeed, since bulk Si absorbs strongly in the visible range, the Si thickness that contributes to the Raman signal is much smaller than the Rayleigh length of our focused laser beam and the assumption of a semi-infinite Si layer is valid. However, bulk Si becomes quasi-transparent in the near-infrared region and a Si thickness on the order of the Rayleigh length contributes to the Raman signal. Therefore assuming that the Raman signal stems from a semi-infinite Si layer would lead to strong overestimation of the Si Raman signal\cite{soubelet_resonance_2016}. 
Finally, in order to obtain a quantity proportional to the square modulus of the Raman susceptibility $\left|{\alpha}\right|^2$ (see comparison between experimental and theoretical values in Figure 2 in the main text), we have also considered the distinct frequencies and occupation numbers of the \epr and \apr phonon modes (see Eq.1 in the main text).

\providecommand{\latin}[1]{#1}
\providecommand*\mcitethebibliography{\thebibliography}
\csname @ifundefined\endcsname{endmcitethebibliography}
  {\let\endmcitethebibliography\endthebibliography}{}


\end{document}